\documentclass[journal]{IEEEtran}

\usepackage{graphicx}
\graphicspath{{figures/}}
\DeclareGraphicsExtensions{.pdf,.jpeg,.png,.jpg}
\usepackage{amsfonts}
\usepackage{algorithmic}
\usepackage{textcomp}
\usepackage{mathtools}
\usepackage{subfigure}
\usepackage{amsmath}
\usepackage{amssymb}
\usepackage{multirow}
\usepackage{multicol}
\usepackage{booktabs}
\usepackage{array}
\usepackage{tabulary}
\usepackage{stfloats}
\usepackage{xcolor}
\usepackage{mathrsfs}
\usepackage{bm}
\usepackage{url}
\usepackage{cite}
\usepackage{gensymb}
\usepackage[ruled]{algorithm2e}
\usepackage{gensymb}
\usepackage[normalem]{ulem}
\usepackage{verbatim}
\usepackage{soul}
\usepackage{mathtools}
\usepackage{booktabs}

\soulregister\cite7
\soulregister\ref7
\soulregister\pageref7

\DeclarePairedDelimiter\abs{\lvert}{\rvert}%
\DeclarePairedDelimiter\norm{\lVert}{\rVert}%

\makeatletter
\let\oldabs\abs
\def\abs{\@ifstar{\oldabs}{\oldabs*}}
\let\oldnorm\norm
\def\norm{\@ifstar{\oldnorm}{\oldnorm*}}
\makeatother

\begin{document}
\title{Microwave Breast Imaging via Neural Networks for Almost Real-time Applications}
\author{Michele~Ambrosanio,~\IEEEmembership{Member,~IEEE,}~Stefano~Franceschini,~\IEEEmembership{Student Member,~IEEE,}\\~Vito~Pascazio,~\IEEEmembership{Senior Member,~IEEE,}~and~Fabio~Baselice.

\thanks{The authors are with the Department of Engineering, Universit\`a degli Studi di Napoli Parthenope, 80143, Napoli, Italy, and with the Consorzio Nazionale Interuniversitario per le Telecomunicazioni (CNIT), viale G.P. Usberti, 43124, Parma, Italy.
E-mail: \{michele.ambrosanio,~stefano.franceschini,~vito.pascazio,~fabio.baselice\}@uniparthenope.it}
}

\maketitle

\begin{abstract}
Conventional breast cancer imaging techniques are nowadays based on the use of ionising radiations or ultrasound waves for the inspection of breast areas. Nevertheless, these conventional techniques present some drawbacks related to patient's safety, processing time and resolution issues. 
In this framework, microwave imaging can represent a valid alternative or a complementary technique compared to other conventional medical imaging modalities since it is safe (using non-ionising radiations), relatively cheap and more comfortable from patient's point of view. 
Unfortunately, it is slow and computationally expensive, which strongly limit its use in clinical scenarios. 
In this paper, an artificial neural network for effective and almost real-time breast imaging is proposed.
First, a realistic breast-like phantom generator was developed for training the network. Subsequently, numerical analyses have been conducted for the optimisation and the performance evaluation of the approach. 
The results seem very promising in terms of recovery performance as well as for the computation burden.
\end{abstract}

\begin{IEEEkeywords}
Microwave tomography, breast imaging, neural networks, artificial intelligence, electromagnetic inverse scattering.
\end{IEEEkeywords}

\section{Introduction}
\label{sec:Intro}

\subsection{Motivation}
\label{sec:into:motivation}
\IEEEPARstart{B}{reast} cancer is one of the commonest types of cancer which affects the women and its early detection is of vital importance for successful treatments \cite{fitzmaurice2015global, siegel2015cancer}. Among the commonest clinical imaging and diagnostic modalities employed for this aim it is worth to mention X-ray mammography, magnetic resonance imaging (MRI), ultrasound scanning and nuclear medicine \cite{jaglan2019breast}. 

X-ray mammography is based on the use of very-high-frequency radiations (10 PHz - 10 EHz) which travels along the scanned area. Thus, it exploits ionizing radiations and it represents an uncomfortable exam due to compression of the breast to carry out the diagnosis. Some other limitations are related to the limited dynamic range, low contrast and grainy image, for which it is difficult to visualise very subtle lesions in women having implants or surgical scars \cite{herranz2012optical}. Another disadvantage is related to its poor spatial resolution and the requirement of large storage place.

Conversely from the use of ionising radiations to perform the investigation, the use of breast MRI is more recommended for women having high risk, representing a valuable alternative to X-ray mammography since it does not involve any ionising radiation exposure. This medical exam produces good spatial resolution images but has some limitations related to the low specificity which results into further tests and biopsies, which drives into higher expenses \cite{sehgal2006review}. Besides being expensive, this imaging modality is not portable, slow in the acquisition process and unsuitable for patients with metallic devices. 

Breast ultrasound, which is based on the use of ultrasonic mechanical waves for non-invasive diagnostics, is used as a follow-up test for abnormalities found by mammogram and provides some guidelines for biopsy inspections. This medical examination produces both qualitative as well as quantitative diagnostic information with a good image quality \cite{kapur2004combination}. Main drawbacks are related to the operator-dependent nature of this exam, to the poor resolution of the image and the low contrast.

Complementary information can be inferred from nuclear medicine, whose functional images are based on the molecular properties of the tissues and on the injected radioactive substance. This imaging modality uses ionising radiation, like in the mammography case, and it is very expensive. Conversely from the other imaging modalities, it can investigate the physiological function of the system, but has limited resolution and slow imaging time \cite{james2014medical}. 

The drawbacks and limitations of the aforementioned diagnostic methodologies have motivated the research community to develop new imaging techniques and modalities to realise an early, reliable and not-expensive diagnostics. In this framework, the use of microwaves for breast imaging and cancer detection has received significant attention, since they might provide better sensitivity and safety due to their non-ionising nature \cite{nikolova2011microwave, aldhaeebi2020review}.

Microwave tomography exploits microwave signals to investigate breast tissues and provide quantitative permittivity and conductivity maps (the dielectric properties) of the imaged tissues. 
This imaging technique offers several advantages over other classical methods since it is not expensive, non-ionizing, and comfortable respect to the treatment \cite{o2018microwave}. These systems provide an evaluation of the electrical properties of the tissues which proved to be a good discriminator between healthy and malignant tissues \cite{lazebnik2007large, fear2002enhancing, conceiccao2016introduction}. 

In particular, microwave systems can provide complementary information on the investigated regions which can be merged with the one deriving from traditional exams to support medical decisions. 
The main issue which limits the adoption of such systems lies in the complicated signal processing and imaging algorithms. As a matter of fact, the scientific research community is focused on the development of more efficient and faster imaging approaches.

\subsection{Prototypes}
\label{sec:into:prototypes}
Due to the potentialities of the microwave imaging modality, several research groups developed and realised different prototypes in recent years \cite{zhurbenko2010design, fedeli2018tomograph, tobon2019design, pagliari2015low, fear2013microwave, porter2013time, santorelli2015time}. These systems all share the feature of employing more antennas arranged around the object of interest with more (virtual or real) transmitters and receivers with the aim of producing an image of the objects located in the investigation region. For this reason, these approaches are also named \textit{tomographic}.
Nowadays, some of these imaging systems have started their active clinical trials, showing the potentialities of this technology as an ally to support medical decisions. 

Initial experimental setups consist of a three-dimensional (3D) prototype for microwave tomography with more transmitters and one receiver which surrounded the imaging domain and were controlled by a motion control system \cite{semenov1999three}. In this case, the object to be imaged was immersed in a deionized water (used as matching medium) in a cylindrical chamber. Other systems followed this setup with an increment in the number of employed antennas \cite{joisel1999microwave}. 
After these preliminary microwave imaging experiences, some clinical prototypes were developed and tested still preserving the cylindrical arrangement of the sensors around the imaging area \cite{meaney2000clinical, meaney2009clinical}, but employing a different kind of antennas immersed in a saline coupling medium to improve signal penetration inside the tissues. After that, several improvements followed in both hardware and software \cite{grzegorczyk2012fast,son2010preclinical, simonov20113d} which yielded to the latest arrangement of the antennas located in a hemispherical shape to make the exam more comfortable on patient's side \cite{klemm2008experimental, preece2016maria, zhurbenko2010design}.

\subsection{Algorithms}
\label{sec:into:algorithms}
In order to obtain the permittivity and conductivity values of the imaged tissues, an inversion algorithm has to be adopted for solving the electromagnetic inverse scattering (EIS) problem which is ill-posed and strongly nonlinear in its general formulation \cite{colton2019inverse}. 
The non-linearity of the considered inverse problem is related to the multiple scattering interactions between points of different tissues and also inside the same tissue, which makes the problem at hand hard to be solved and affected by false solutions.
These two challenging issues have stimulated different research groups to develop several methods in order to provide tomographic qualitative as well as quantitative images of the biological tissues under test \cite{colton2019inverse, bertero1998introduction}. In this framework, two approaches can be identified: the tomographic methods, which aim at producing a full map of the biological tissues in terms of permittivity and conductivity or in terms of labels which are uniquely related to the considered tissues (i.e., segmentation and classification maps), and the radar-based techniques, which aims at identifying a pathology within a region and not to provide an image of the region under test \cite{aldhaeebi2020review, benny2020overview}.

Due to the need of quantitative analyses for diagnostic purposes, the capability of tomographic approaches of providing maps of the investigated biological tissues makes them very attractive for clinical applications. Thus, several imaging strategies have been developed across the past thirty years to solve this problem.
With regard to the tomographic approaches, three main categories can be identified: qualitative, approximated and quantitative methods \cite{pastorino2010microwave}. 

Qualitative methods aim at solving an inverse obstacle problem \cite{salman2005microwave, bolomey1982microwave, colton2003linear} by processing the scattered field and providing an estimation of the total tissue extension in the breast, but not its characterisation (i.e., the type of tissue). 
Approximated methods exploit some approximations of the scattering phenomena in order to allow easy implementation and to keep the computational complexity low. Although they are quite fast, they suffer for some limitations related to the adopted approximated model, as in the case of the well-known Born and Kirchhoff approximations \cite{cui2001inverse}. Another relevant linearised approach proposed in recent years for biomedical imaging purposes is based on the virtual experiments framework \cite{crocco2012linear}, which is based on the combination of real and ``virtual'' experiments that propose a linear approximation of the EIS problem in the case of non-weak scattering regime. Furthermore, higher-order Born approximations can be exploited to reconstruct the conductivity function of the dielectric tissues under test \cite{cui2004low}.

In order to overcome the limited retrieving performance related to the aforementioned classes, quantitative approaches can be exploited \cite{belkebir1996newton}. With regard to these methods, the class of retrievable objects becomes wider at the expense of higher computational burden and processing time. Different iterative approaches can be employed to solve this problem and to face the issues related to non-linearity which may drive into false solutions due to the presence of local minima. Due to the high computational complexity of these methods some local minimisation approaches are adopted and thus the choice of the initial step is paramount \cite{nocedal2006numerical}.
However, when the non-linearity of the problem at hand is very strong, then it might be beneficial to use global optimization to avoid local minima \cite{pastorino2002global, pastorino2010microwave}. Unfortunately, their complexity grows exponentially with the number of unknowns and this makes their use very hard to be applied for realistic and/or real-time applications. Moreover, due to non-linearity, the inversion procedure may be more sensitive to modelling errors and uncertainties on the scenario.

\subsection{Machine learning for quantitative microwave imaging}
\label{sec:into:ml}
In the framework of inverse problems, fast and reliable non-linear approaches are desirable for addressing the imaging problem in the biomedical area of breast cancer diagnostics. Among the most recent methodologies, artificial neural networks represent a useful and flexible tool for quantitative imaging. As a matter of fact, neural networks and artificial intelligence proved to perform well in the field of computer vision, image processing and classification. 
First methodologies based on artificial neural networks were applied to extract some general information about the geometric and electromagnetic properties of the scatterers and tissues at hand \cite{caorsi1999electromagnetic, rekanos2002neural}. Most of these first attempts to face the imaging problem via neural networks used a few spatial as well as electromagnetic parameters to represent the scatterers.

Recently, most of the literature has focused on the use of deep convolutional neural netoworks (CNNs) for solving the inverse problem \cite{xu2020deep, li2018deepnis, wei2018deep, sanghvi2019embedding, guo2019supervised, massa2019dnns, aggarwal2018modl}. Neural networks with regression features have provided very impressive results on EIS problems. 
The majority of these articles do not propose a direct inversion scheme, i.e. the approach does not allow to move directly from the data collected at receivers to an estimate of the profile, but they usually perform a super-resolution of the recovery starting from a raw image obtained via other conventional approaches. 
One of the most adopted techniques consists in the training of a U-net architecture \cite{lucas2018using} for obtaining quantitative recoveries via preliminary manipulations, e.g. approximated models and a-priori information to move from the data (i.e., the scattered field samples) to contrast/induced currents approximations \cite{wei2018deep, li2018deepnis}.  
In order to solve the EIS problem with high contrast, a contrast-source-based neural network combined with traditional subspace-based optimisation method and CNNs might be employed \cite{sanghvi2019embedding}. Furthermore, neural networks can be also employed as regularisation strategies in conventional inversion approaches \cite{ashtari2010using, khoshdel2020full} as well as to obtain super-resolved reconstructions \cite{shah2017super}.

Conversely from the contemporary scientific literature which focuses on the use of CNNs, in this work we focused on artificial neural networks (ANNs) based on multilayer perceptrons. This kind of networks allows to implement a direct inversion scheme from the scattered field samples to directly retrieve a quantitative map of the dielectric features of breast profiles in a fast, efficient way.
Despite the ease in network design, a critical issue lies in the choice of large enough data set for training the network, which proves to be of vital importance for the estimation of the links strength between nodes \cite{lucas2018using}. Thus, the main bottleneck of this kind of approaches is related to the required computational burden to train the network. Nevertheless, after the initial training, then a direct mapping between data and unknowns can be obtained, producing reliable images in a considerably fast inversion procedure. 

Inspired by the universal approximation theorem (UAT) \cite{hornik1990universal}, which states that any arbitrary non-linear function can be approximated via a proper fully-connected neural network with a large number of neurons in its hidden layers under some mild assumptions, in this manuscript we propose an ANN architecture for the almost real-time \textit{quantitative} imaging of female breast dielectric properties. The motivation in choosing such an architecture is supported by the need of defining a general approach for the retrieval of whatever non-linear profile, as supported by the UAT. 
As a matter of fact, in an ANN architecture all the inputs contribute to every single output, resulting more suitable for this kind of applications rather than CNNs, the latter being the default choice when dealing with highly-structured modalities, such as images or video.

The outline of the paper is as follows. In Section \ref{sec:problem} an overview of the mathematics involved in electromagnetic inverse scattering (EIS) problem is recapped. In Section \ref{sec:methodology}, an overview of the proposed fully-connected ANN-based approach is reported with a focus on the data set generation dealing with a realistic breast-like phantom generator for the database population. Finally, some results on synthetic breast phantoms are shown in Section \ref{sec:result} and a comparison with conventional nonlinear EIS approaches is proposed. Some conclusions are drawn at the end of the article.     

\section{Problem Statement}
\label{sec:problem}
In the following, a simplified two-dimensional geometry is considered. The background medium is supposed to be homogeneous with complex permittivity $\epsilon_b$ and with magnetic permeability $\mu_0 = 4\pi \cdot 10^{-7}$ H/m. 
The reason for such an assumption is related to the fact that biological tissues are characterised by a constant value of the magnetic permeability while a certain variability of the complex permittivity can be observed.
The antennas are located along a measurement curve $\Gamma$ which surrounds the imaging domain $\Omega$. The targets located inside this domain are illuminated via transverse-magnetic electric fields generated by z-oriented current wires located on $\Gamma$.
\begin{figure}[t]
   \centerline{ \includegraphics[draft=false, trim=0cm 0cm 0cm 0cm, width=0.3\textwidth]{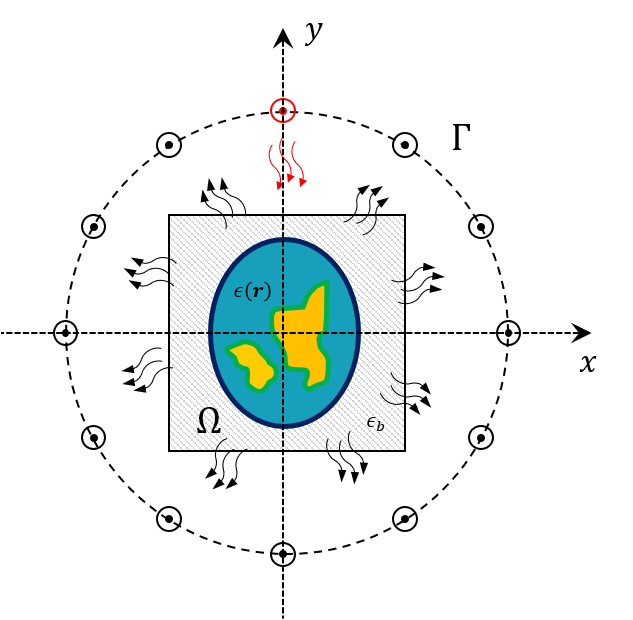}}
    \caption{Sketch of the 2D geometry. The antennas are located on the measurement line $\Gamma$ and arranged in a multiview-multistaic fashion.} 
    \label{fig:geometry_2D}
\end{figure}
A sketch of the geometry is reported in Fig. \ref{fig:geometry_2D}. The scattering phenomena depend on the contrast between the dielectric properties of the background medium and those ones of the target. The aim of the proposed method is to provide the breast tissues properties maps by solving an EIS problem, i.e. their relative complex permittivity $\epsilon\left(\bm{r}\right) = \epsilon'\left(\bm{r}\right)-j \frac{\sigma\left(\bm{r}\right)}{\omega\epsilon_0}$, with $\epsilon'\left(\bm{r}\right)$ and $\sigma\left(\bm{r}\right)$ being the relative permittivity and conductivity maps, respectively. 
Thus, the electromagnetic scattering equations ruling these phenomena can be written as \cite{colton2019inverse}: 
    \begin{align}
    \label{eq:ES_equations_data}
        E_s& = \mathcal{A}_e\left(\epsilon', \sigma, E_t\right) + n\,, \\
    \label{eq:ES_equations_domain}
        E_t& = E_i + \mathcal{A}_i\left(\epsilon', \sigma, E_t\right)\,,
    \end{align}
in which the dependence on the operating frequency and background dielectric features have been implied. $E_i$ and $E_t$ represent the incident and total electric field inside the imaging domain $\Omega$, respectively, while $E_s$ represents scattered field at receivers locations on $\Gamma$. The quantities $\mathcal{A}_i$ and $\mathcal{A}_e$ are the radiating operators which depend on the dielectric properties of background medium $\epsilon_b$, and $n$ is the noise which affects the collected data, here assumed to be additive, white and Gaussian (AWGN).  

As regards the imaging applications, the considered framework can be dealt with as an EIS problem consisting in the retrieval of a quantitative estimate of the unknown relative permittivity $\epsilon'$ and conductivity $\sigma$ functions inside $\Omega$ from the scattered field samples measured on $\Gamma$. As previously stated in Subsection \ref{sec:into:algorithms}, such a problem is both nonlinear and ill-posed \cite{isernia1997nonlinear, bucci1997electromagnetic, bertero1998introduction}, thus
finding a solution to this problem is not trivial and requires to face high computational burden and time-consuming approaches. Nevertheless, reliable and fast algorithms able to provide dielectric properties maps of the tissues under test are desirable for early diagnosis in the biomedical field. 
In this framework, conventional nonlinear approaches can achieve good recovery performance \cite{ambrosanio2018multithreshold, bevacqua2019method, mojabi2009enhancement, shea2010contrast, rubaek2007nonlinear} at the expense of high computational burden, which implies no real time applications. Moreover, considerable a-priori information and data pre-processing is required to obtain reliable recoveries.   

In this framework, artificial neural networks (ANNs) can be of interest and represent a very attractive alternative for almost real-time applications with more accurate reconstructions. In this case, the challenge is related to the network design and the data set generation for the training step which considerably impacts on the recovery performance.    
\begin{figure}[!t]
    \centering
    \includegraphics[draft=false,  width=0.45\textwidth]{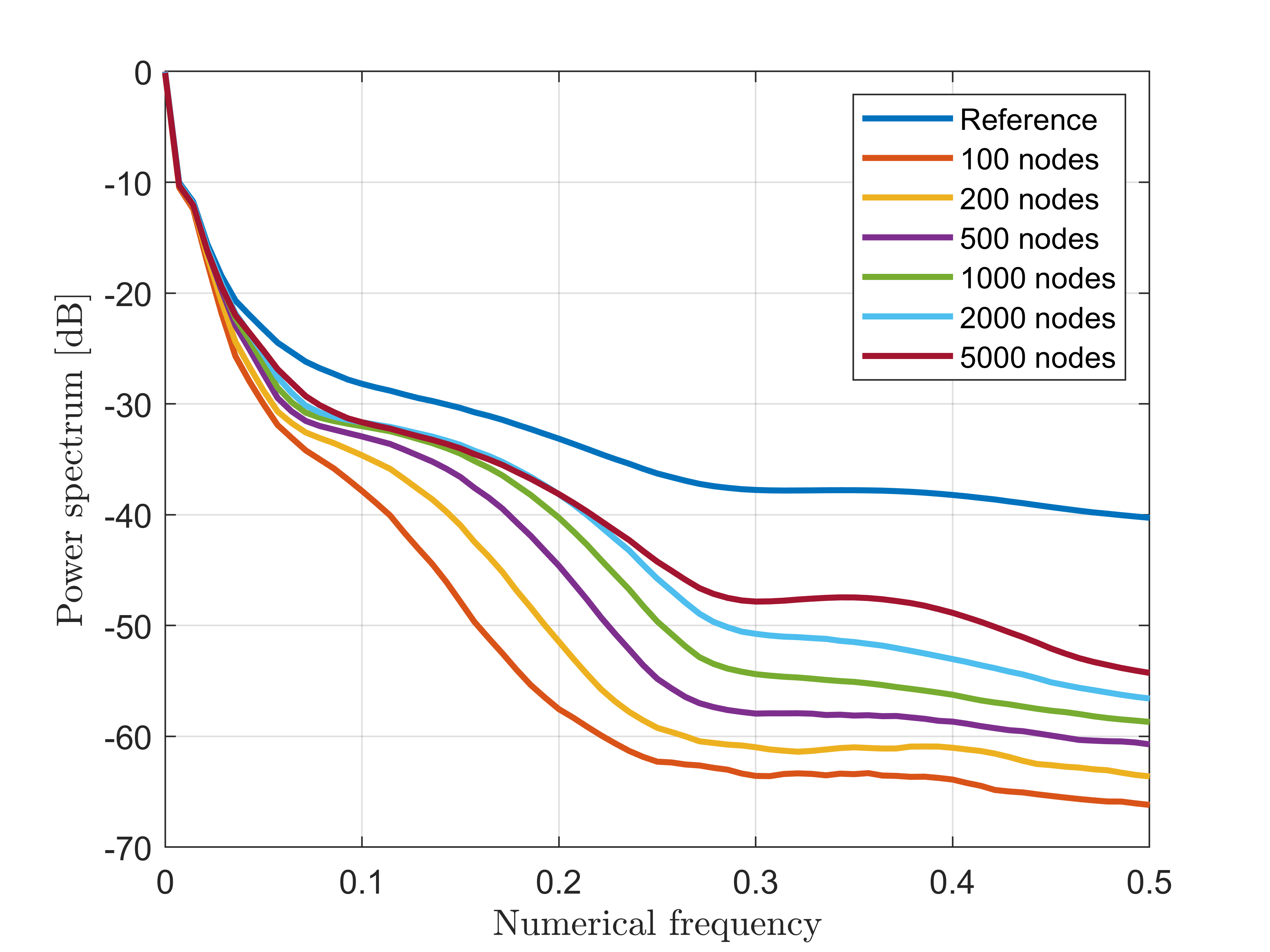}
    \caption{Averaged spectrum $\mathcal{S}(\nu)$ for six different three-layer architectures as a function of nodes number per each hidden layer.}
    \label{fig:mean_spect}
\end{figure}
\begin{figure}[!t]
    \centering
    \includegraphics[draft=false,  width=0.45\textwidth]{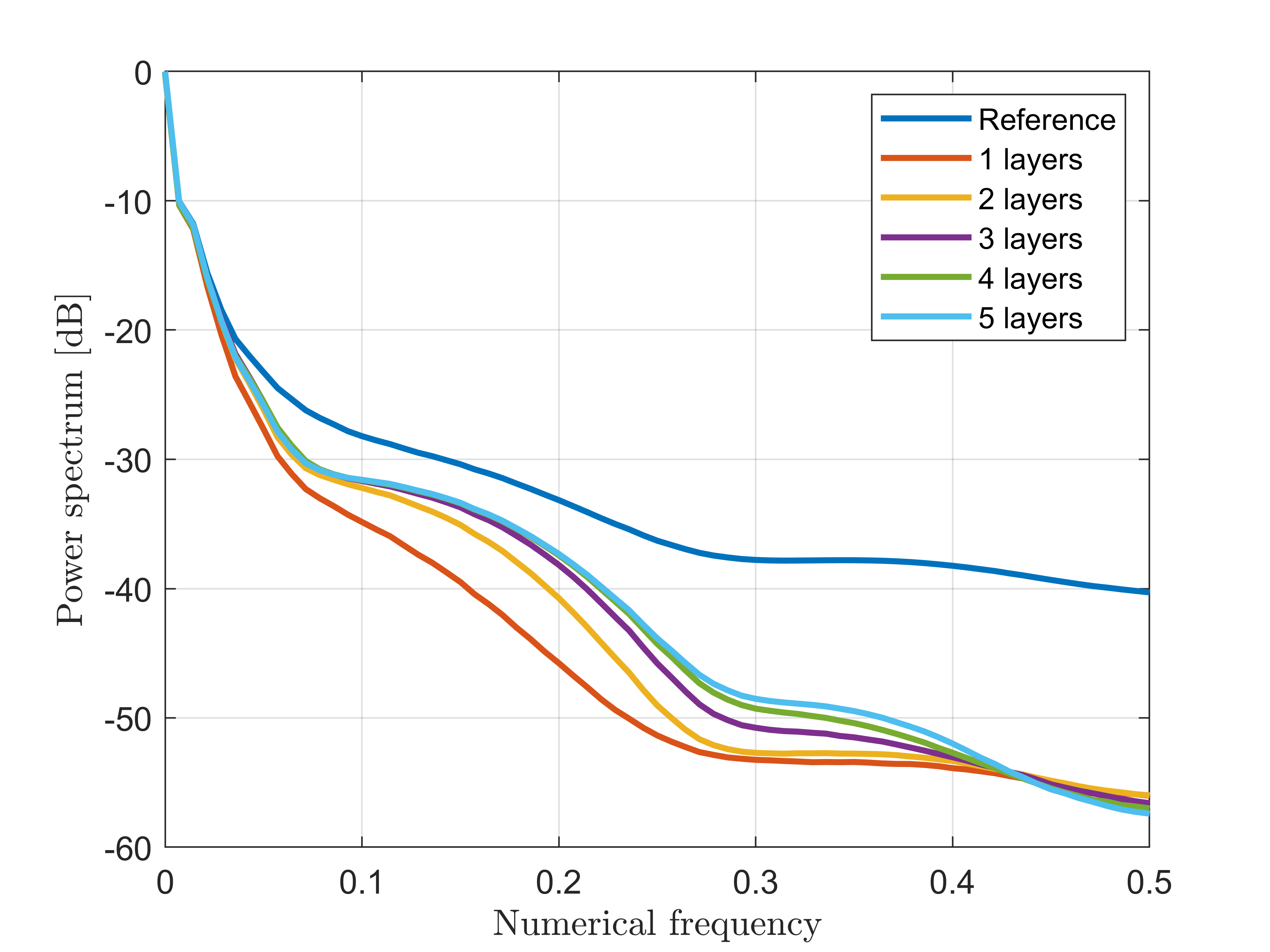}
    \caption{Averaged spectrum $\mathcal{S}(\nu)$ for five different architectures with two-thousand nodes per layer as a function of the number of hidden layers.}
    \label{fig:mean_spect_layers}
\end{figure}

\section{Methodology}
\label{sec:methodology}
In order to define the proposed approach for solving the inverse problem at hand, two main aspects have to be described: the characteristics of the database adopted for the ANN training and the layout of the implemented neural network.

\subsection{Training Database}
\label{sec:methodology:phantom_gen}
In order to perform quantitative inversion via neural networks in a reliable way, it is fundamental to properly model the specific scenario under test and to have large data sets of realistic breast profiles. 
Some research groups focused on the generation of these data sets to be used by the scientific community \cite{lazebnik2007large, lazebnik2007large2}. Unfortunately, in most cases the available population is very limited or not flexible enough to be exploited in the framework of neural networks training, since they are limited to a specific measurement configuration and/or in a certain frequency range.
In order to overcome these limitations and with the aim of providing a useful simulation tool for the testing of breast microwave imaging algorithms, a numerical realistic breast phantom generator is developed to populate the database. After that, the corresponding scattering matrices are calculated exploiting a forward solver based on the method of moments (MoM). These matrices, together with the generated breast profiles, represent the data adopted for the proposed neural network approach.

The first step of the generator consists in the creation of an elliptical-shape phantom. 
The dimensions of the ellipse axes vary uniformly in the range $[6.5, 12]$ cm.
The centre of the ellipse is randomly set inside a circle of radius 1 cm and located in the middle of the imaging domain. The orientation of the ellipse and its thickness, emulating the skin layer, are also randomly selected in the range $[0, 2\pi]$ and $[1.5, 2.5]$ mm, respectively. 

The geometry of breast inner tissues is generated by exploiting a stochastic 2D multi-fractal random field generator \cite{schertzer1989nonlinear}.
The obtained map is segmented into three classes, related to the \textit{fibro-glandular}, the \textit{transitional} and the \textit{adipose} biological tissues.
According to \cite{lazebnik2007large}, we chose to simulate four breast classes characterised by different tissues percentage, as reported in Table \ref{tab:breast_classes}.

%
\begin{table}[]
\caption{Breast Phantom Classification}
\begin{tabular}{@{}clccc@{}}
\toprule
\multicolumn{1}{c}{\textbf{Class}} & \multicolumn{1}{c}{\textbf{Description}} & \multicolumn{3}{c}{\textbf{Percentage of tissue (\%)}} \\ \midrule
    &                               & Fibro-glandular & Transitional & Adipose \\
I   & Mostly Adipose                & 5 - 20          & 5 - 15       & 65 - 90 \\
II &  Scattered Fibro               & 20 - 30         & 10 - 20      & 50 - 70 \\
III &  Dense                        & 30 - 40         & 15 - 20      & 40 - 55 \\
IV & Very Dense                     & 40 - 65         & 20 - 25      & 10 - 40 \\ \bottomrule
\end{tabular}
\label{tab:breast_classes}
\end{table}

In order to generate realistic values of the tissues dielectric features, they are generated according to the statistical distributions estimated from the database described in \cite{lazebnik2007large}. 
Finally, spatial correlation among neighbouring pixels has been added to the data.
This procedure is repeated to populate the database composed of 120,000 profiles (30,000 per each breast class). For each profile (discretised into $108\times108$ pixels), the scattering matrix has been computed  via a fast Fourier transform-conjugate gradient (FFT-CG) forward solver based on the MoM \cite{isernia1997nonlinear} assuming a multiview-multistatic system with transmitters and receivers located in 30 angular equally-spaced locations on a measurement circle of radius 12 cm. The transmitting signal is a line source at a fixed frequency which impinges on an imaging domain of size $15 \times 15$ cm\textsuperscript{2} discretized according to Richmond's rule \cite{richmond1965scattering}. 

The database has been partitioned in order to use the 85\% of the profiles for the training phase, the 10\% for the validation phase and the remaining 5\% for the testing phase.
\begin{figure*}[t]
   \centerline{ 
   \begin{tabular}{cccc}
        {\includegraphics[draft=false, trim=2.6cm 1cm 2cm 0cm, clip, height=3.7cm]{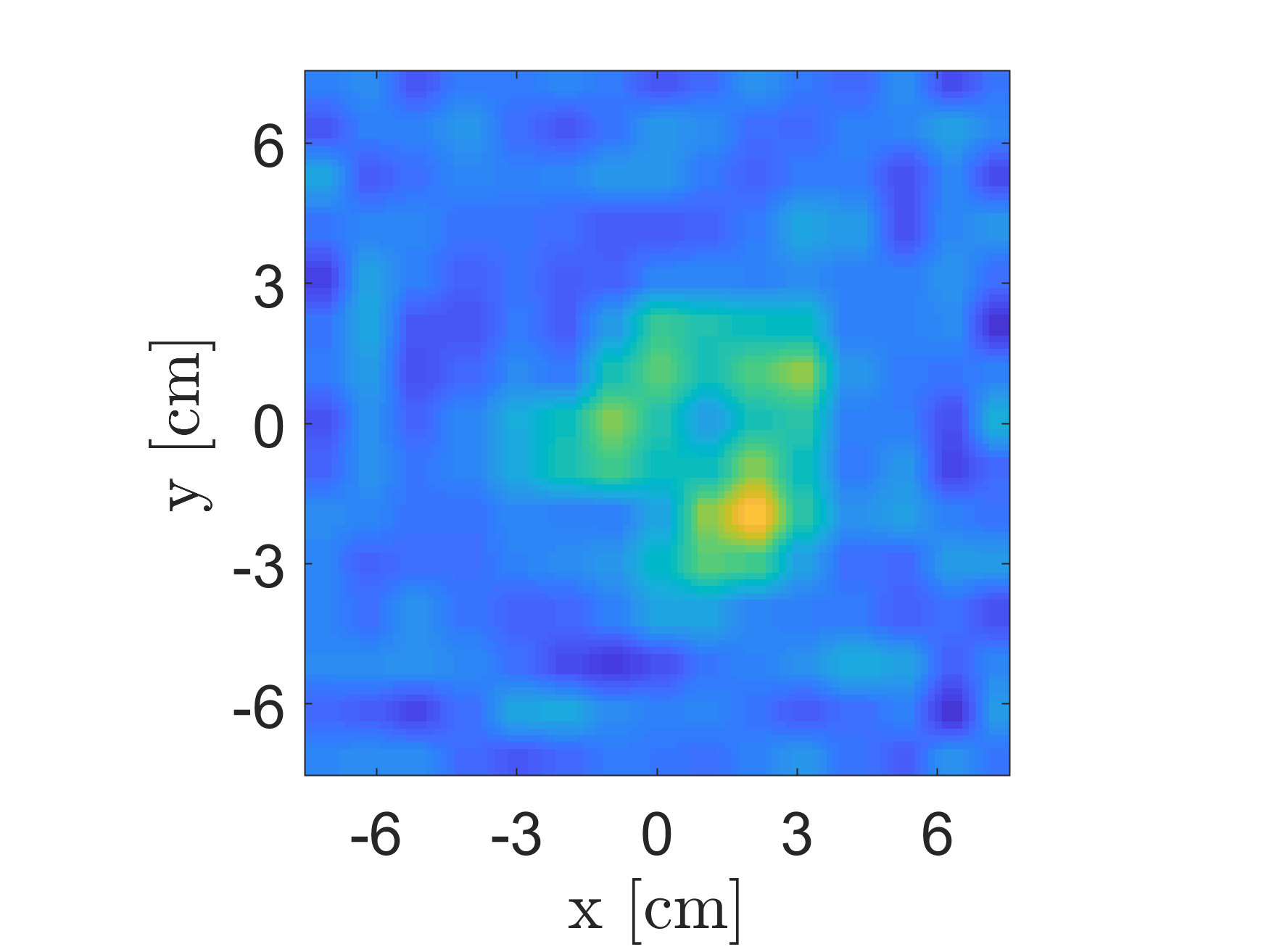}} & 
        {\includegraphics[draft=false, trim=2.6cm 1cm 2cm 0cm, clip, height=3.7cm]{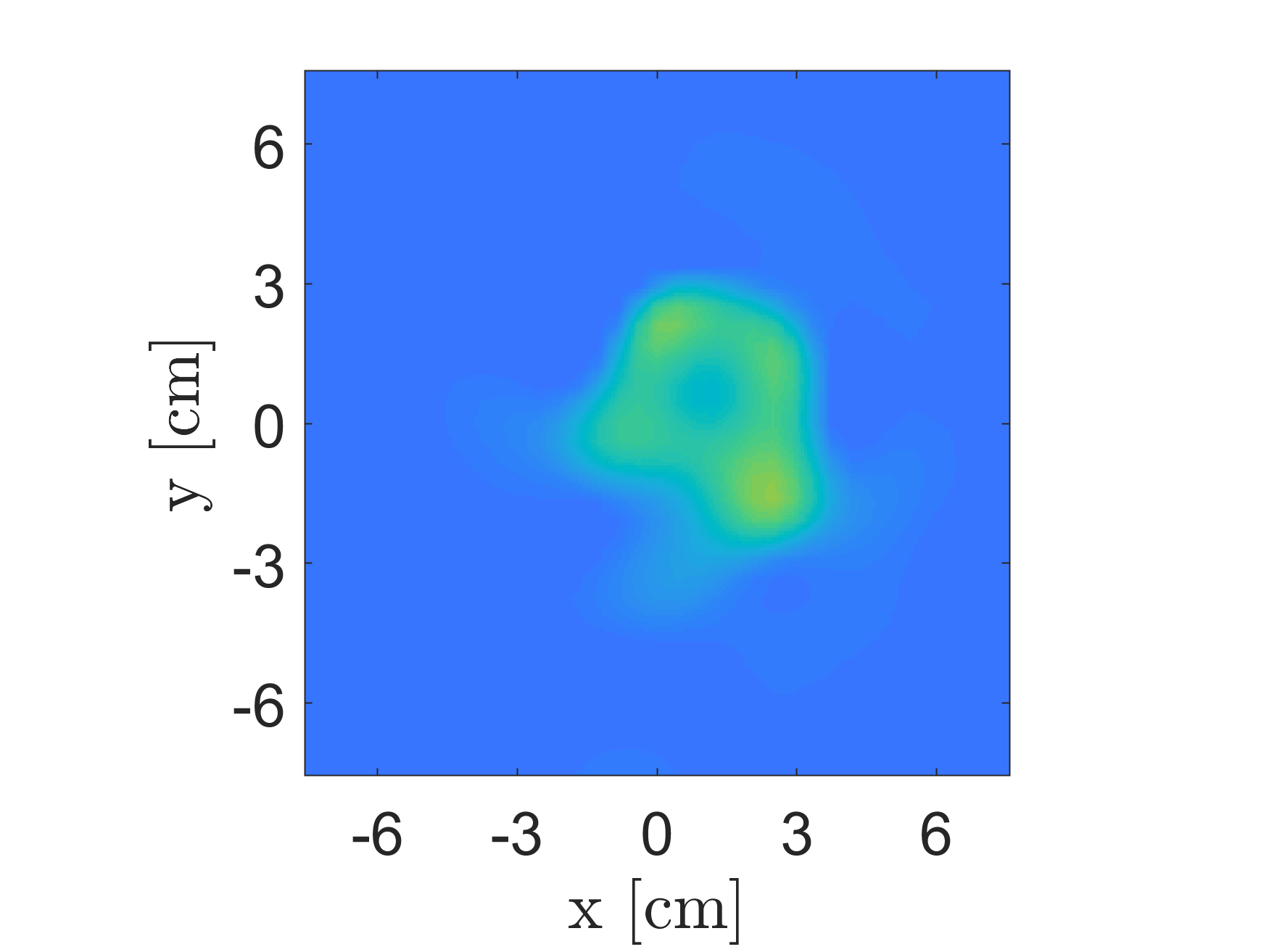}} &
        {\includegraphics[draft=false, trim=2.6cm 1cm 2cm 0cm, clip, height=3.7cm]{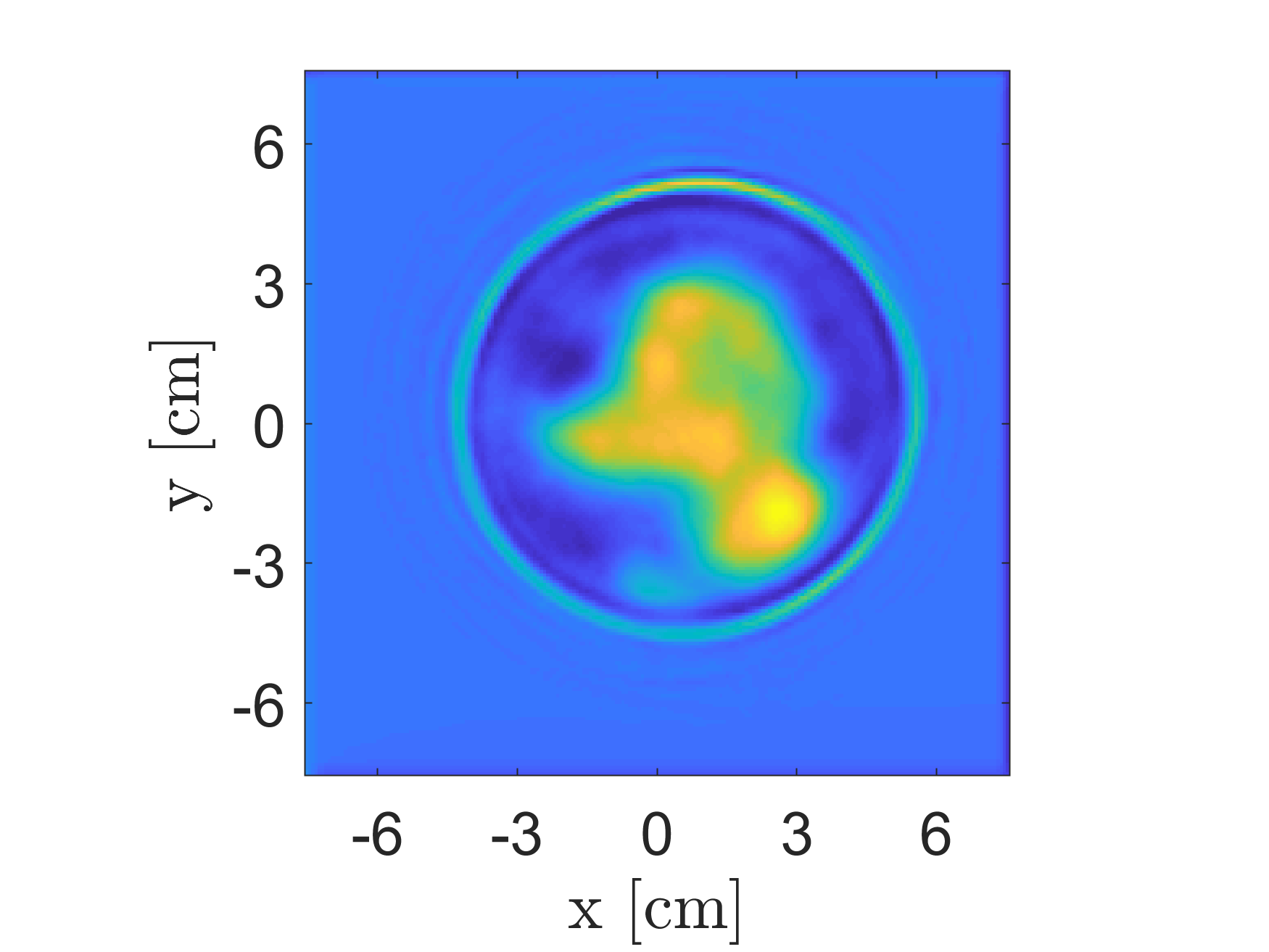}} &
        {\includegraphics[draft=false, trim=2cm 1cm 2cm 0cm, clip, height=3.7cm]{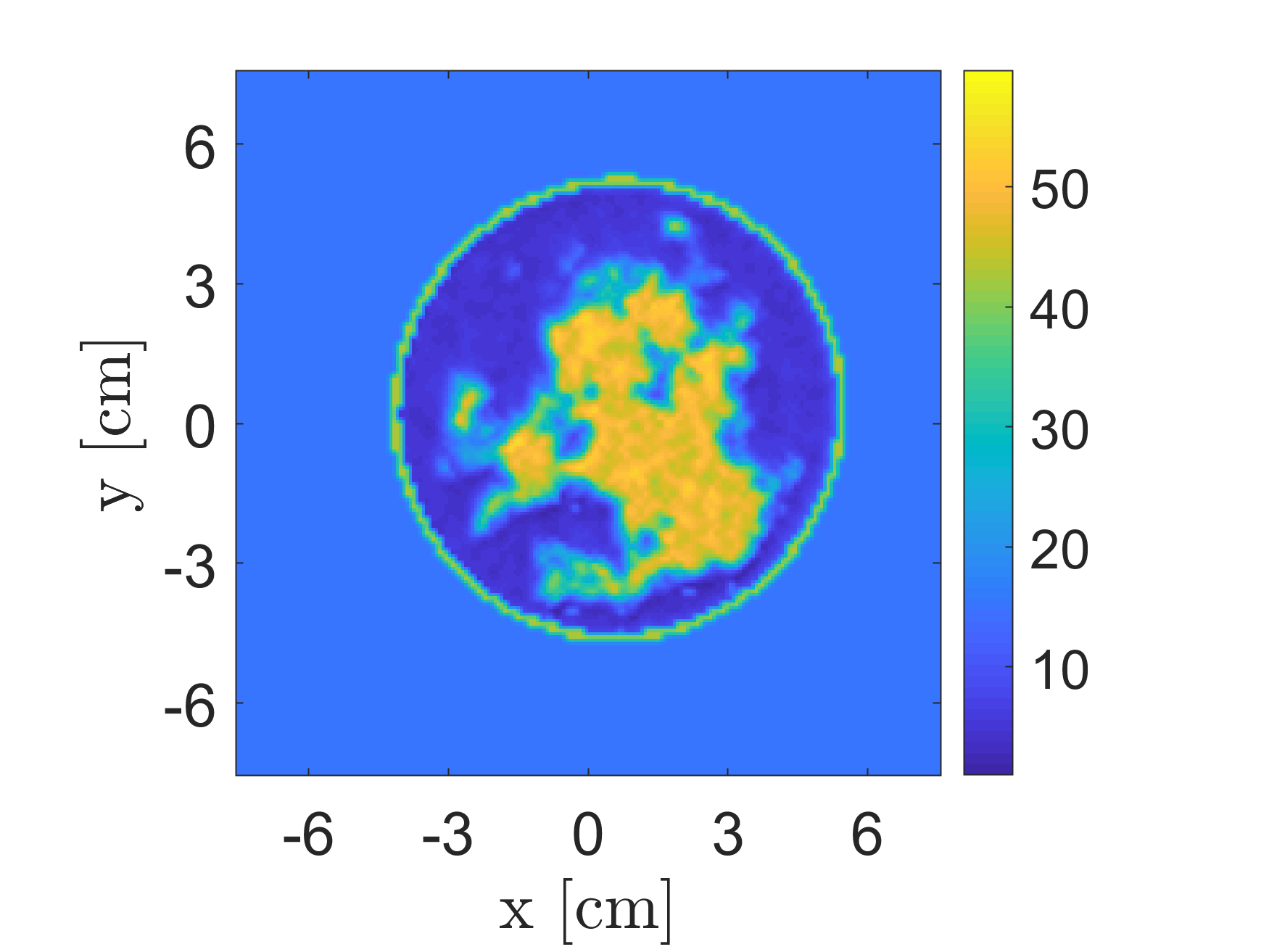}}  \\
 (a) DBIM & (b) CSI & (c) ANN & (d) Reference \\
        {\includegraphics[draft=false, trim=2.6cm 1cm 2cm 0cm, clip, height=3.7cm]{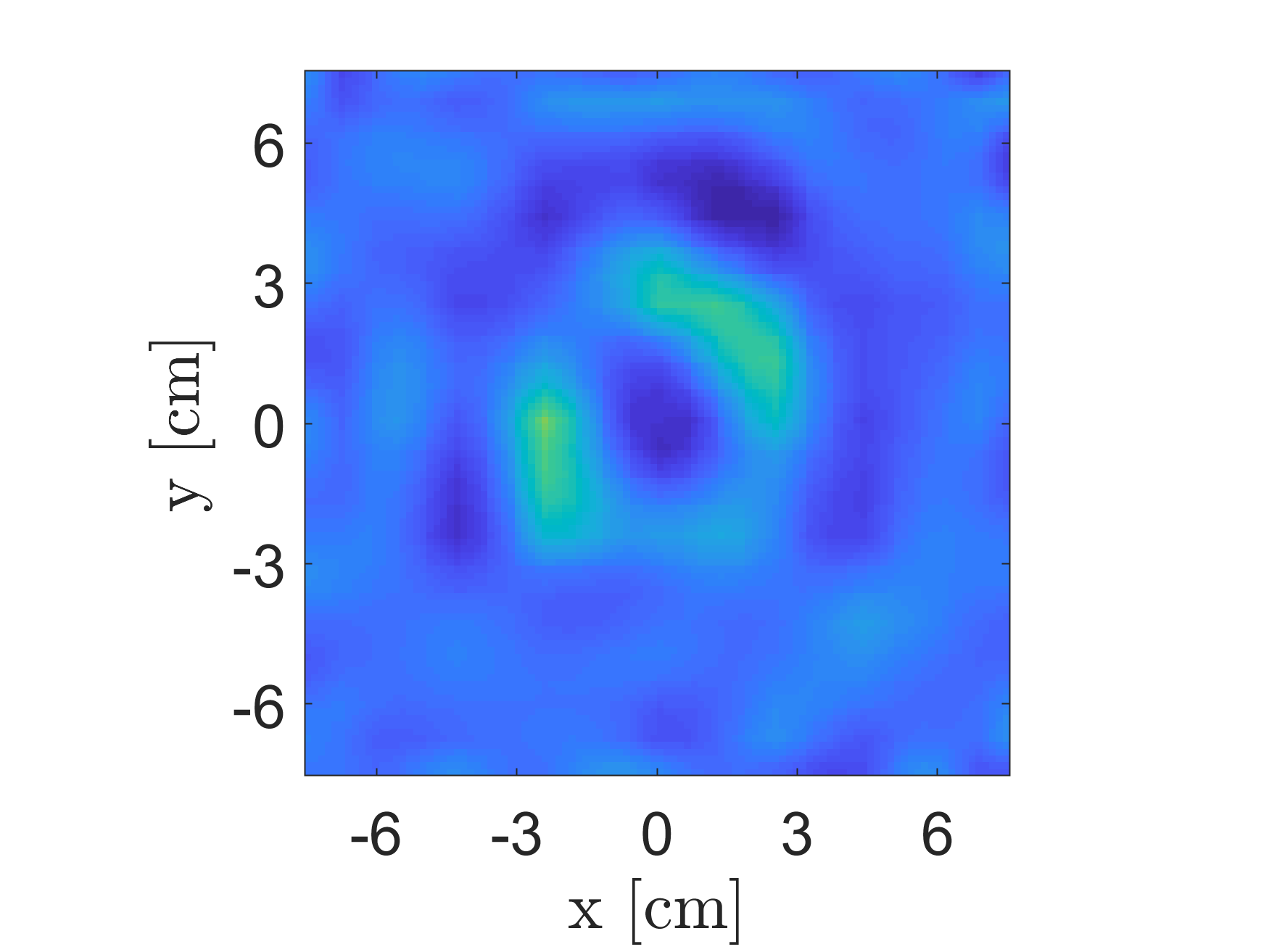}} & 
        {\includegraphics[draft=false, trim=2.6cm 1cm 2cm 0cm, clip, height=3.7cm]{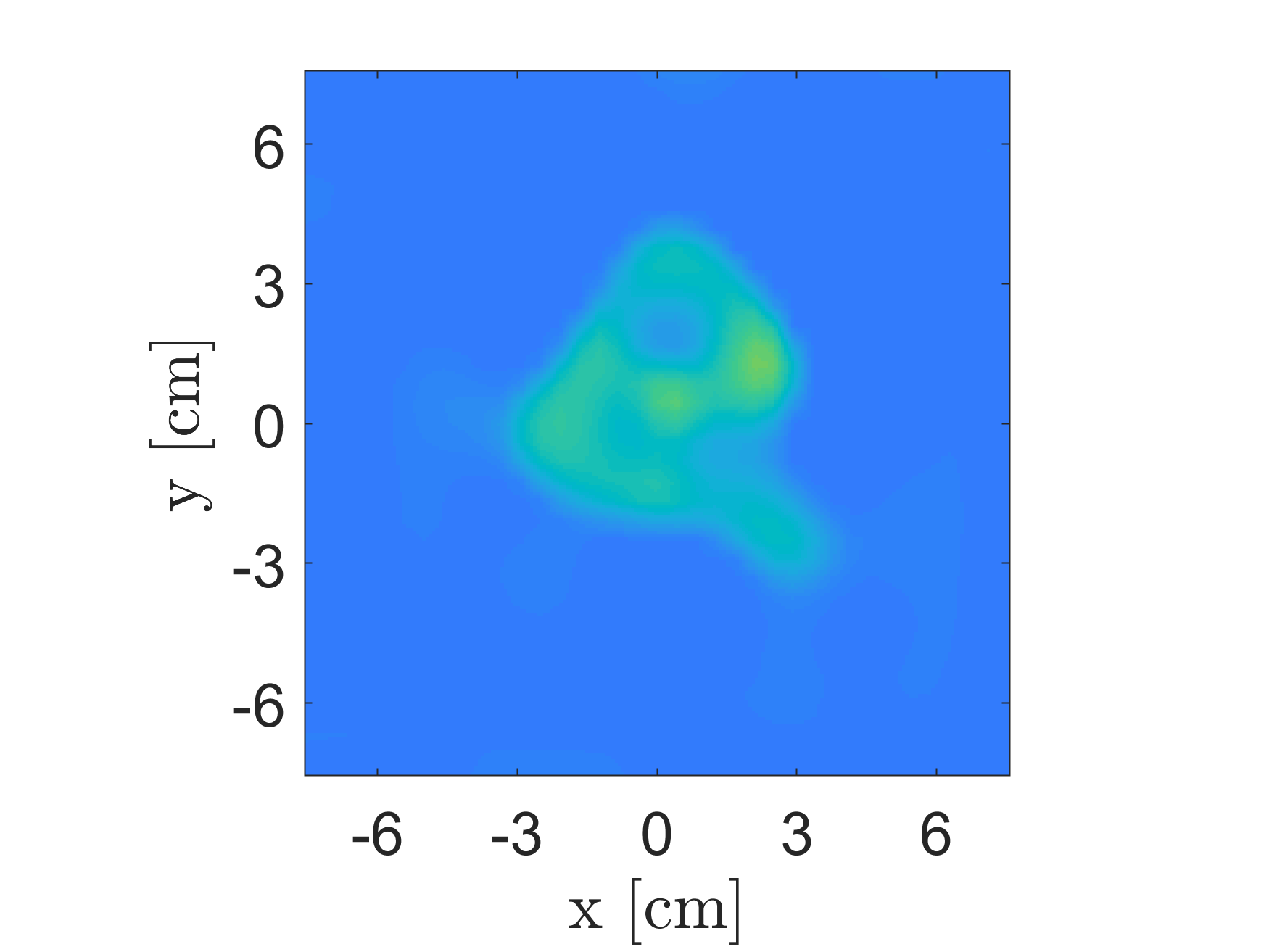}} & 
        {\includegraphics[draft=false, trim=2.6cm 1cm 2cm 0cm, clip, height=3.7cm]{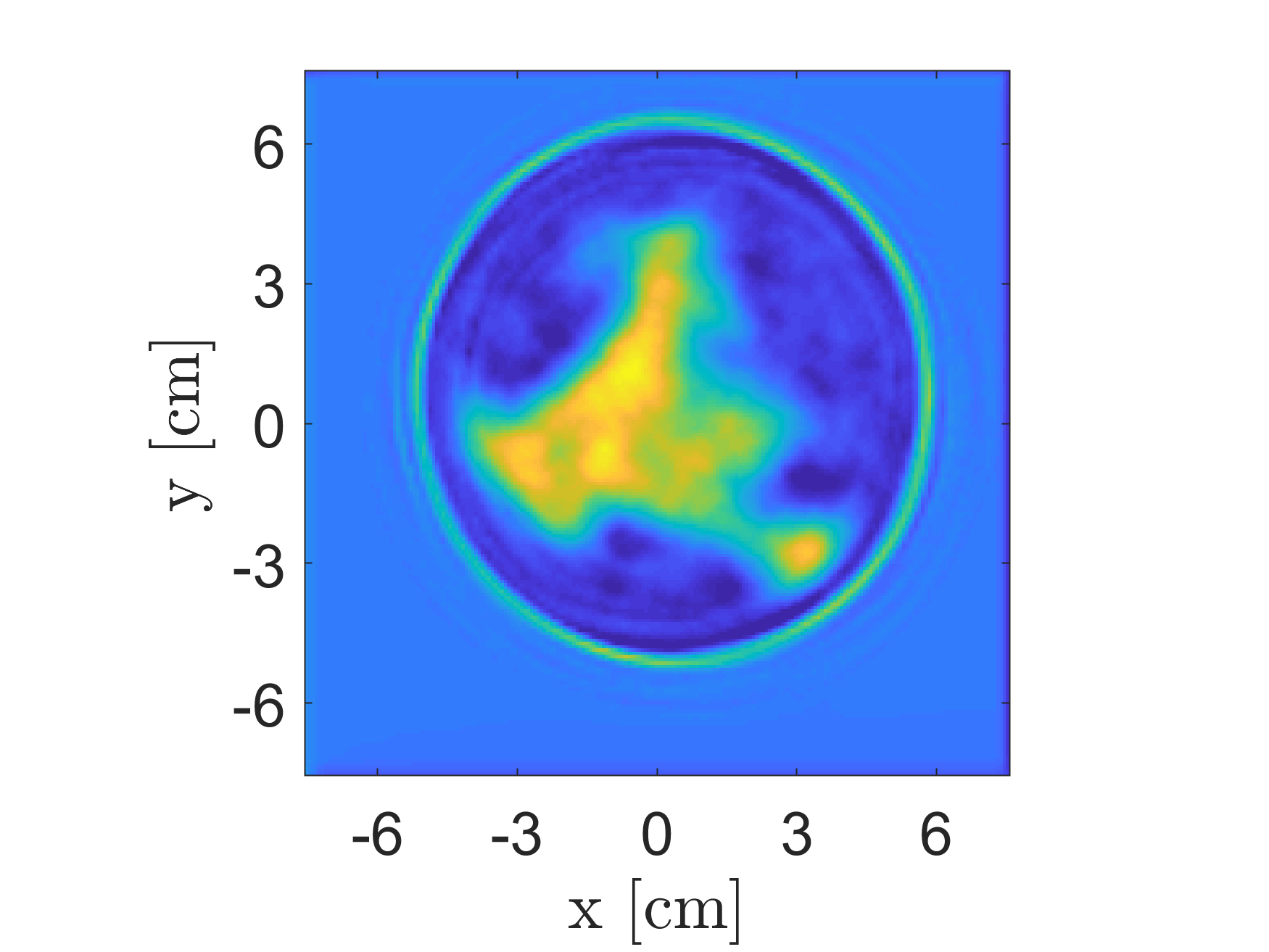}} &
        {\includegraphics[draft=false, trim=2cm 1cm 2cm 0cm, clip, height=3.7cm]{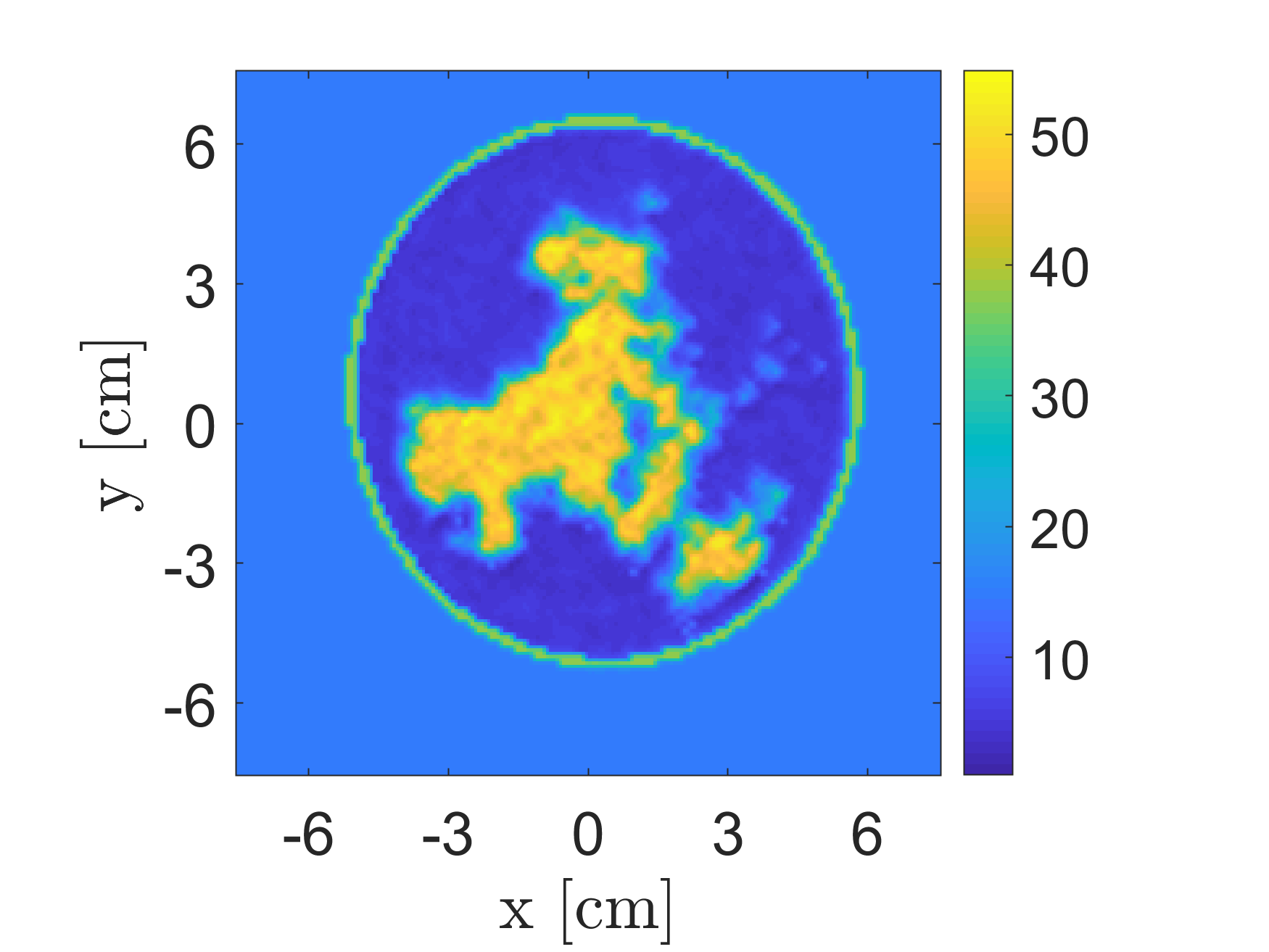}}  \\
 (e) DBIM & (f) CSI & (g) ANN & (h) Reference \\
        {\includegraphics[draft=false, trim=2.6cm 1cm 2cm 0cm, clip, height=3.7cm]{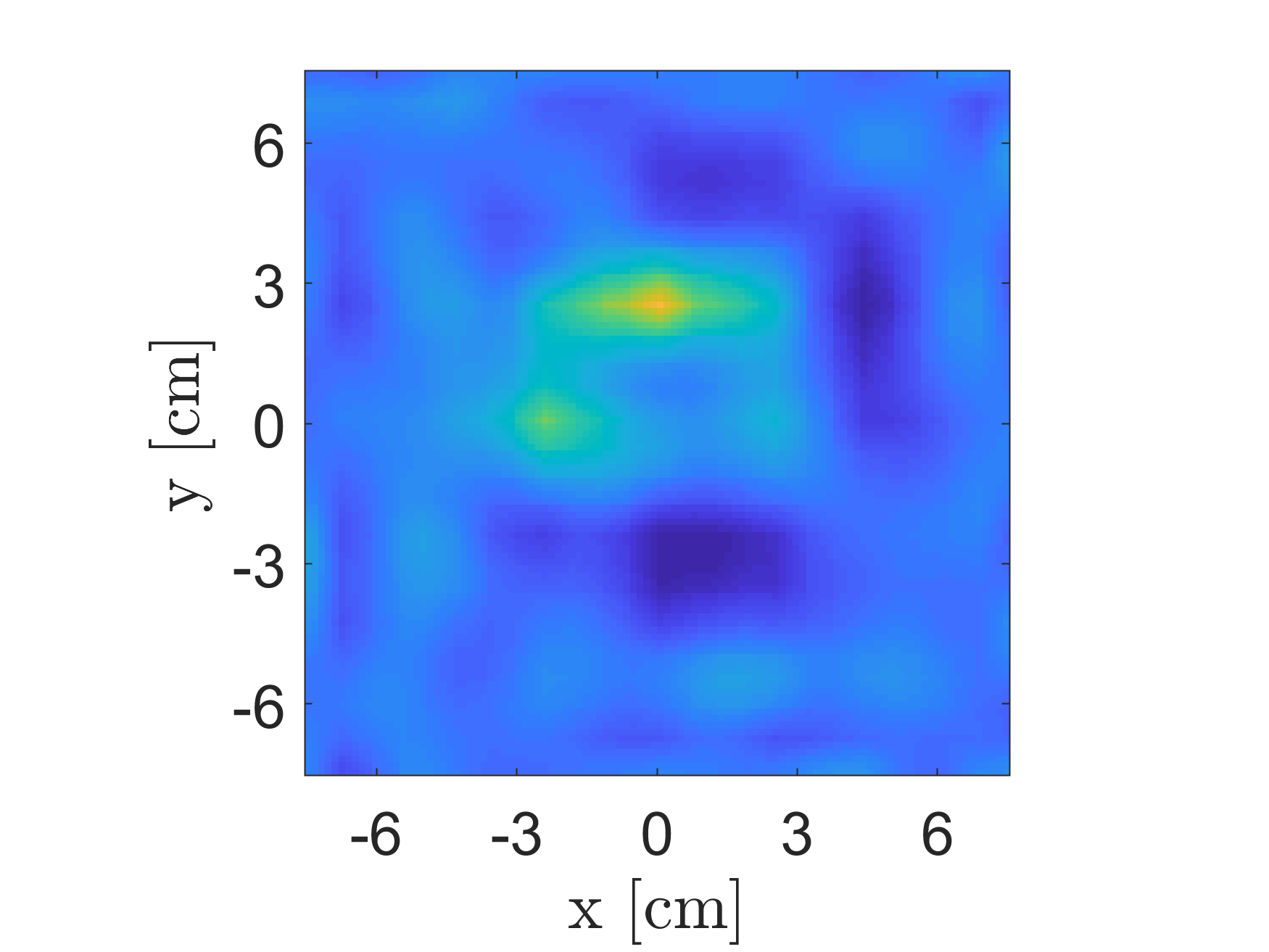}} & 
        {\includegraphics[draft=false, trim=2.6cm 1cm 2cm 0cm, clip, height=3.7cm]{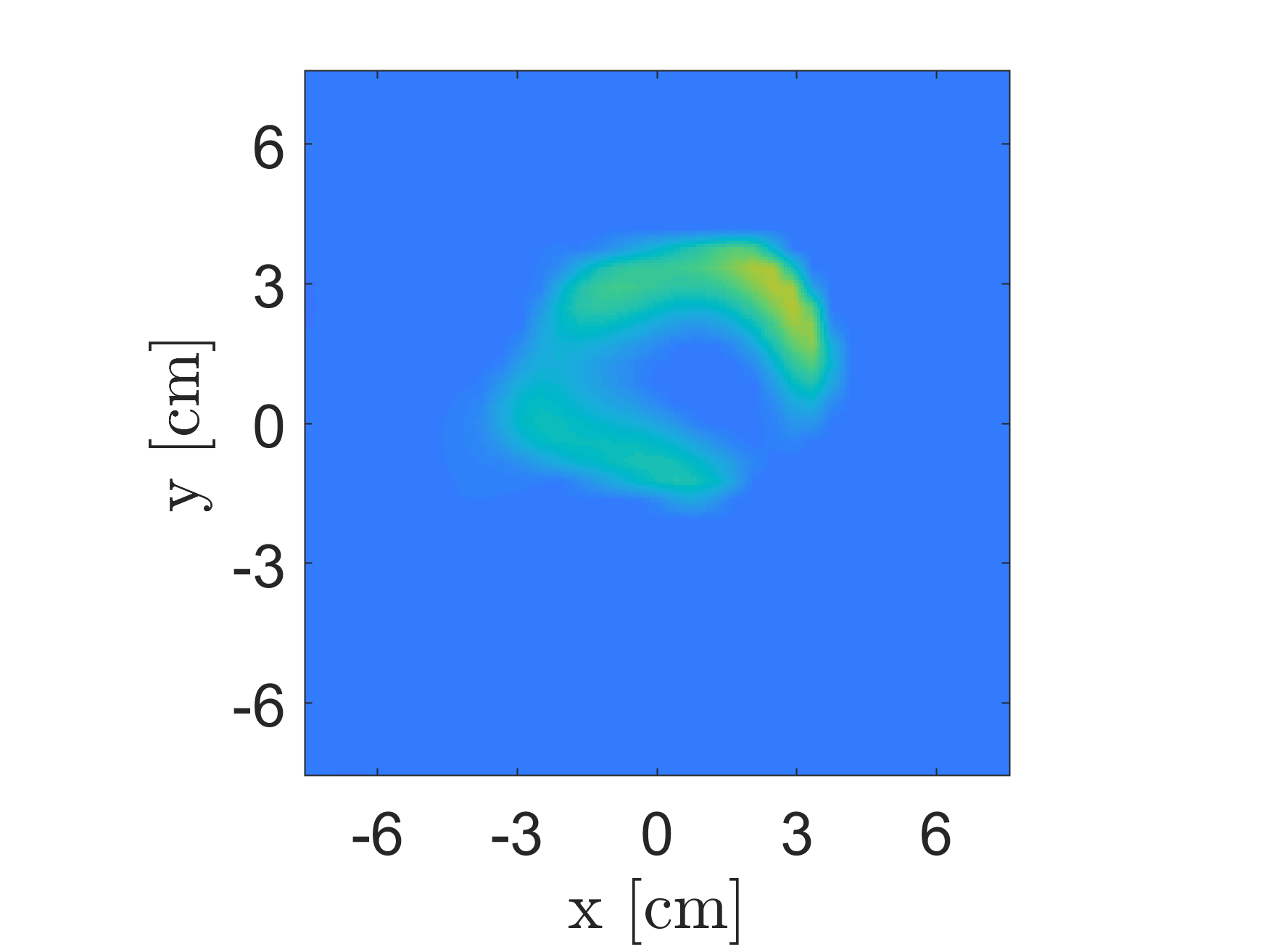}} & 
        {\includegraphics[draft=false, trim=2.6cm 1cm 2cm 0cm, clip, height=3.7cm]{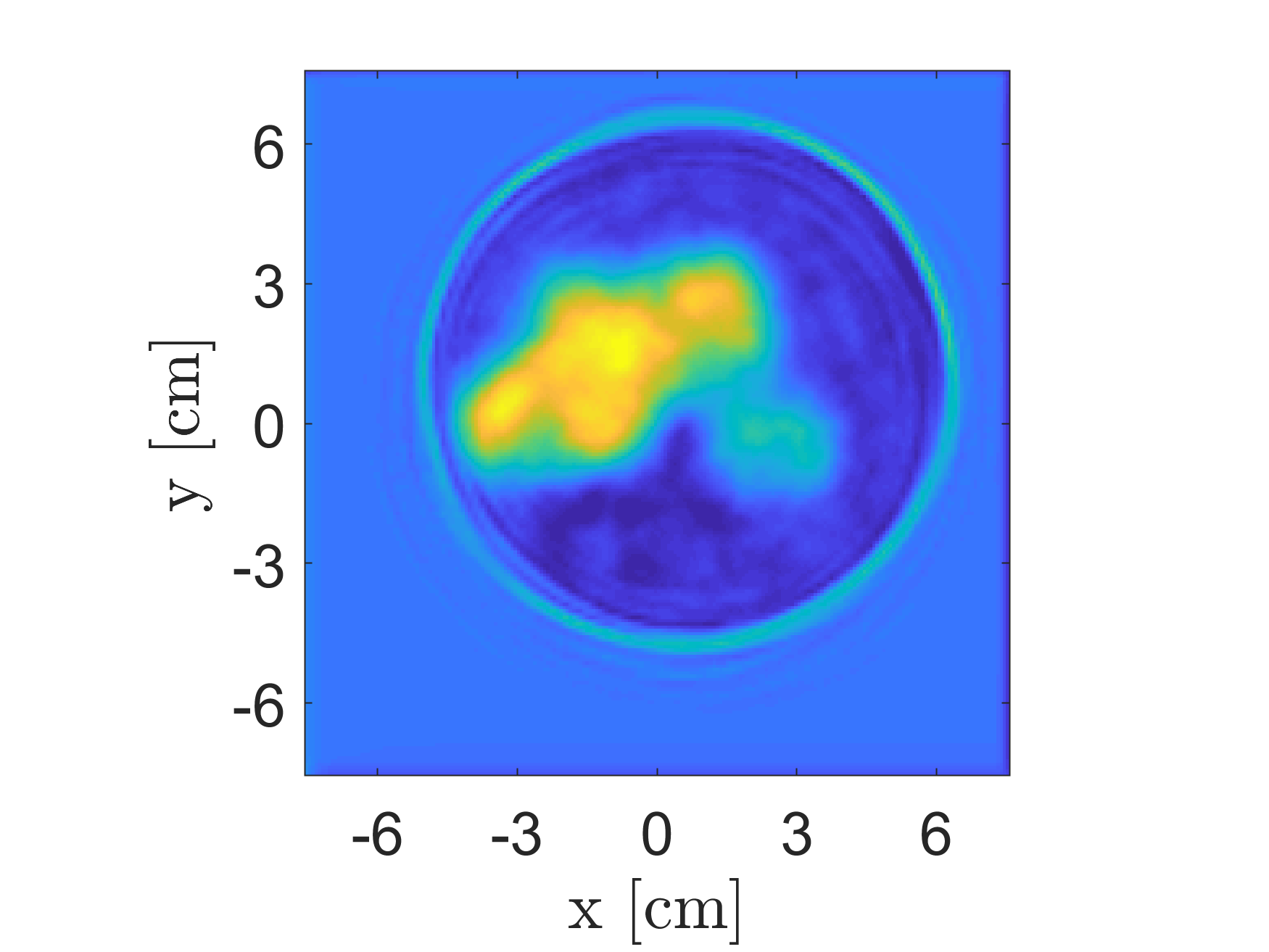}} &
        {\includegraphics[draft=false, trim=2cm 1cm 2cm 0cm, clip, height=3.7cm]{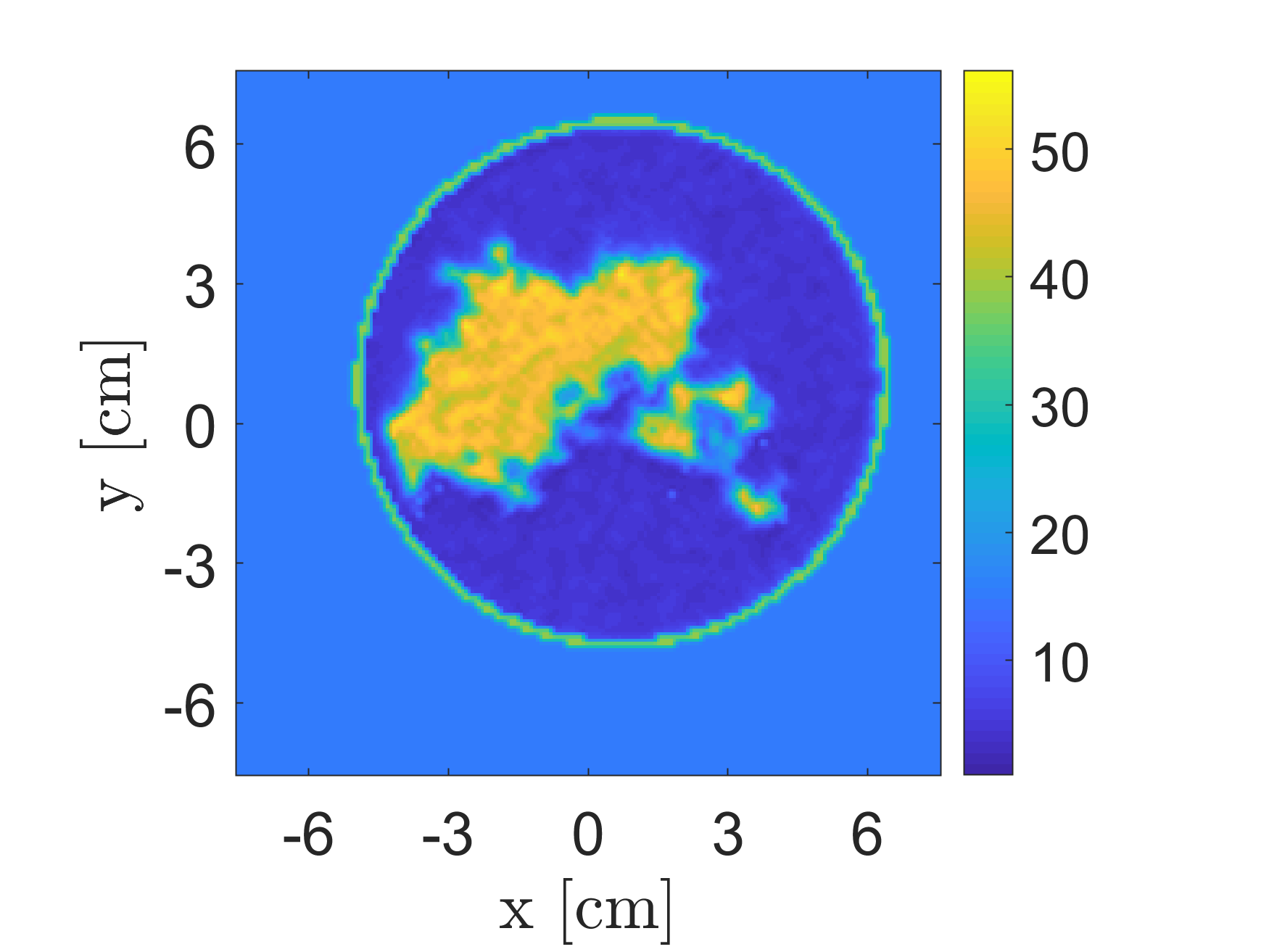}}  \\
(i) DBIM & (j) CSI & (k) ANN & (l) Reference \\
   \end{tabular}}
   \caption{Comparison between recoveries related to three different breast profiles. Retrieved relative permittivity maps for DBIM (a), (e), (i), CSI (b), (f), (j), and ANN-based approach (c), (g), (k). Reference permitivity maps are reported in (d), (h), (l). Both the x and y axes are in centimetre units.}  
   \label{fig:net_comparisons_re}
\end{figure*}
\begin{figure*}[t]
   \centerline{ 
   \begin{tabular}{cccc}
        {\includegraphics[draft=false, trim=2.6cm 1cm 2cm 0cm, clip, height=3.7cm]{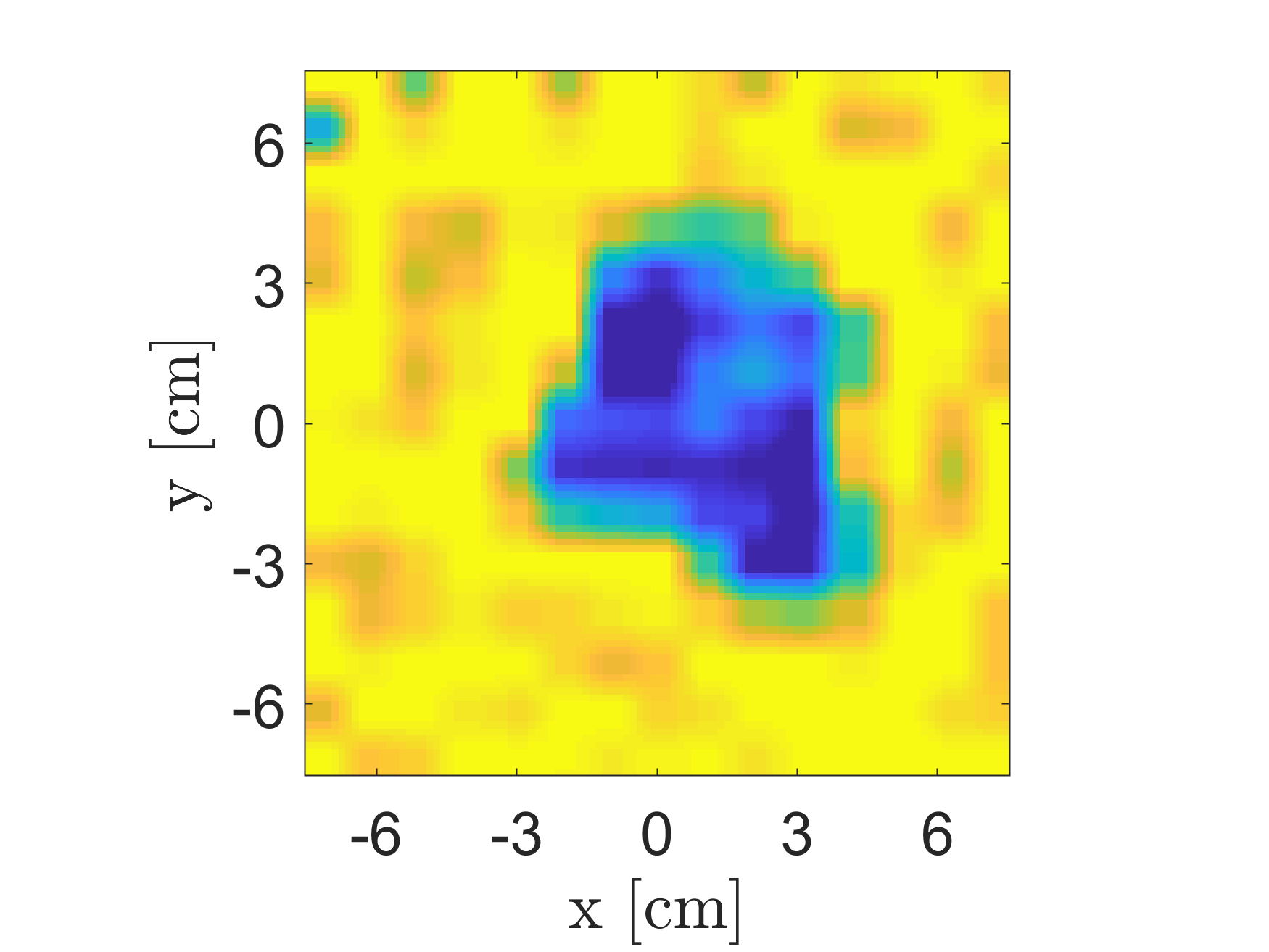}} & 
        {\includegraphics[draft=false, trim=2.6cm 1cm 2cm 0cm, clip, height=3.7cm]{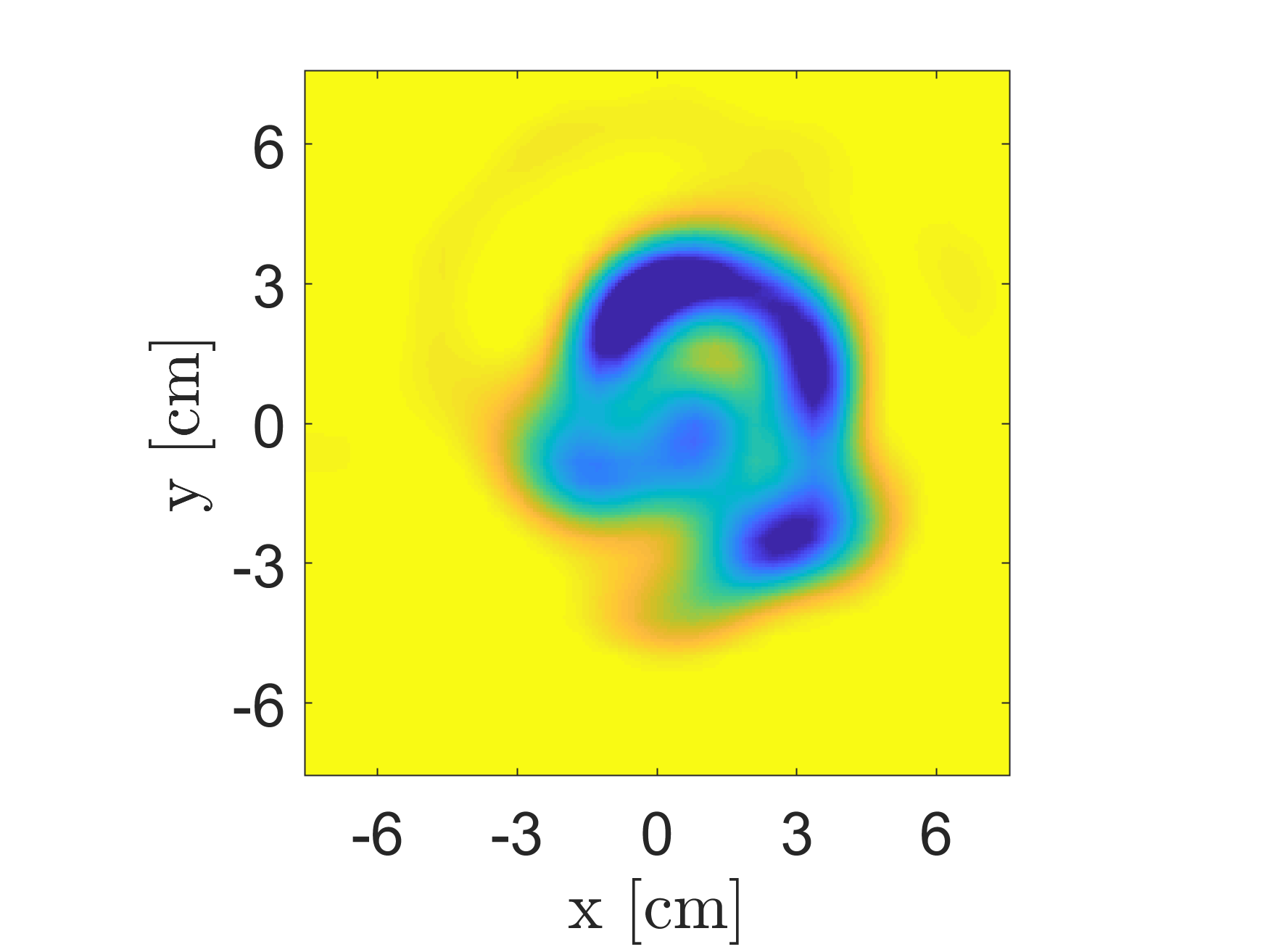}} &
        {\includegraphics[draft=false, trim=2.6cm 1cm 2cm 0cm, clip, height=3.7cm]{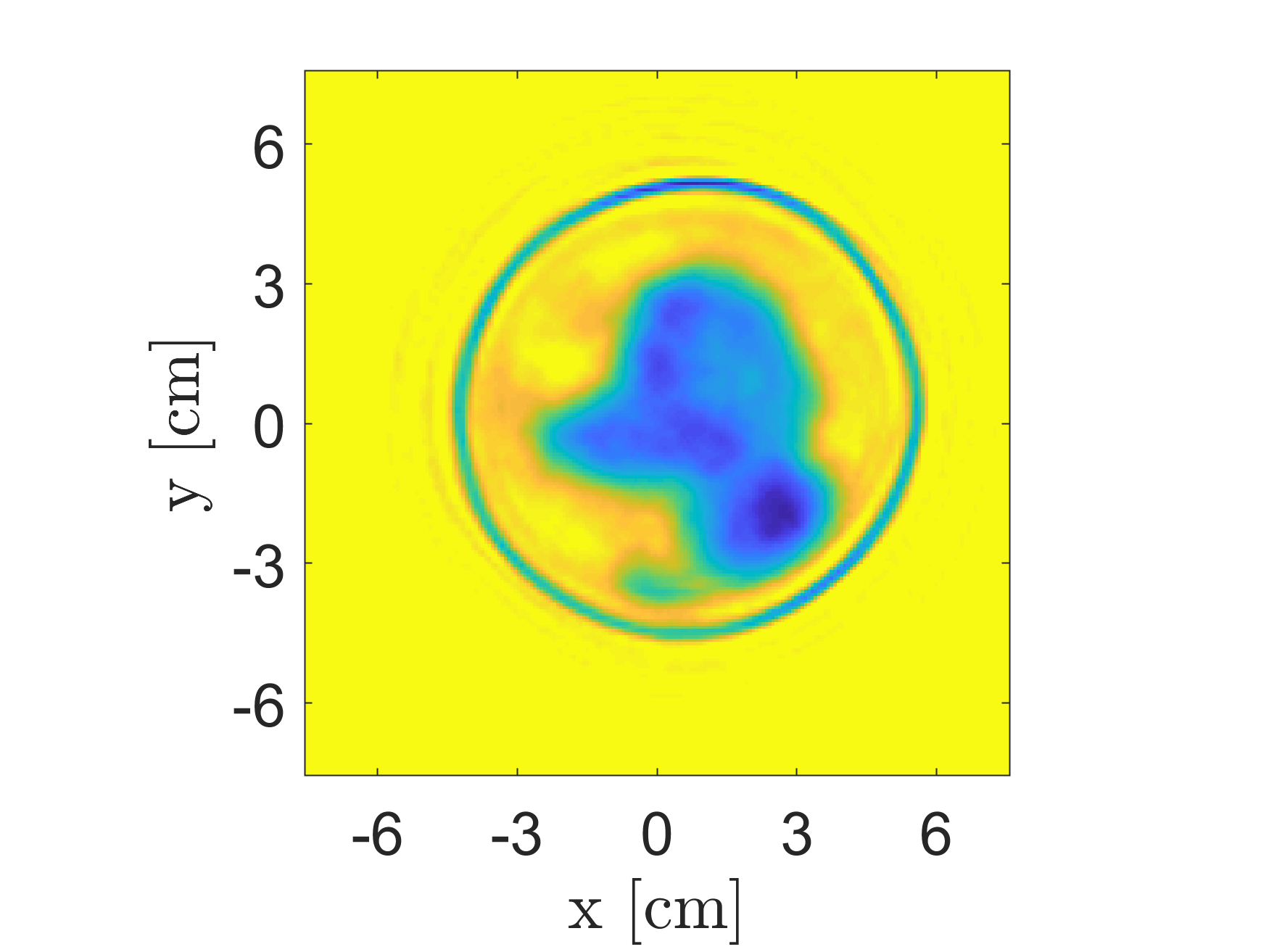}} &
        {\includegraphics[draft=false, trim=2cm 1cm 1.8cm 0cm, clip, height=3.7cm]{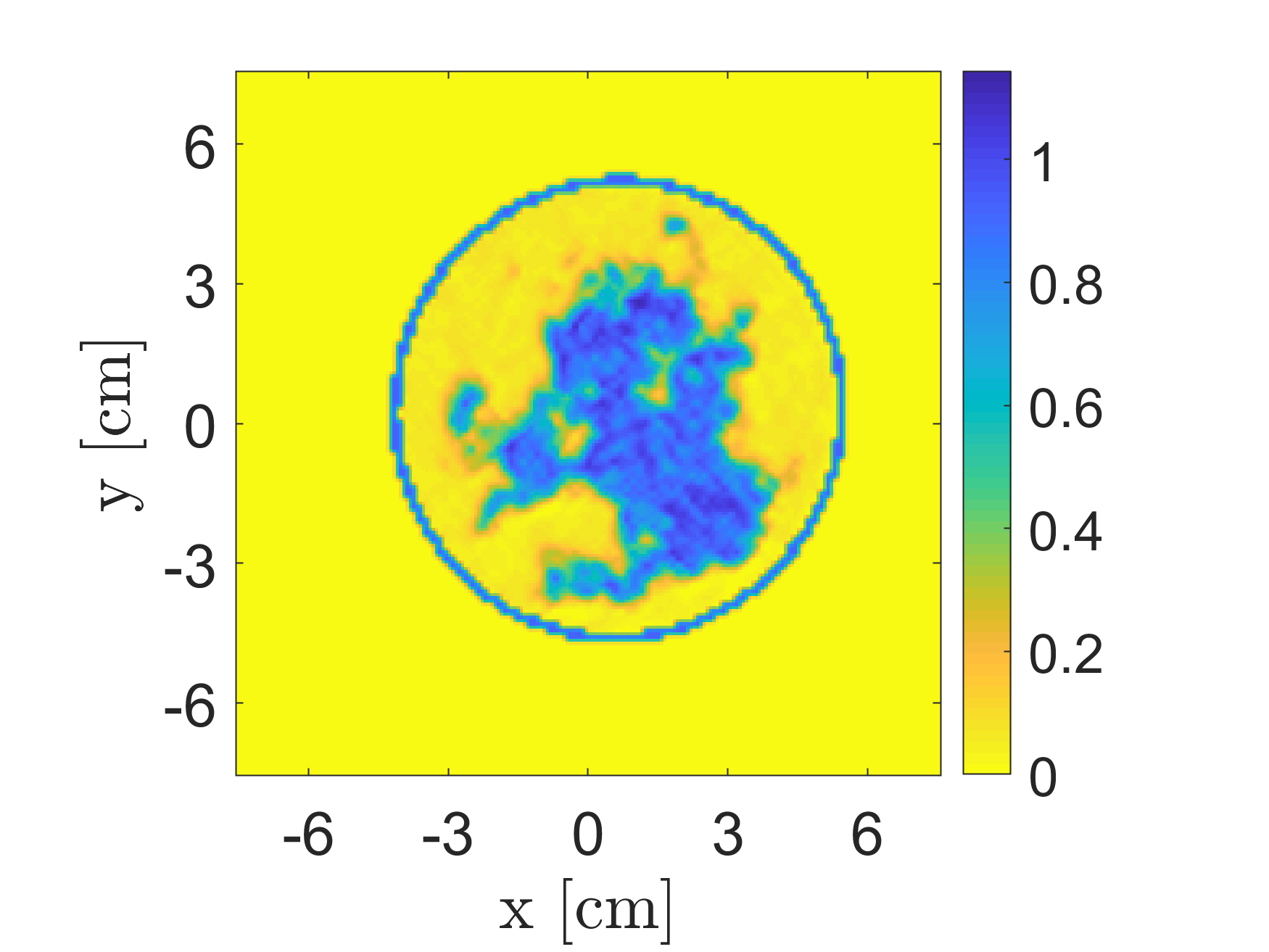}}  \\
 (a) DBIM & (b) CSI & (c) ANN & (d) Reference \\
        {\includegraphics[draft=false, trim=2.6cm 1cm 2cm 0cm, clip, height=3.7cm]{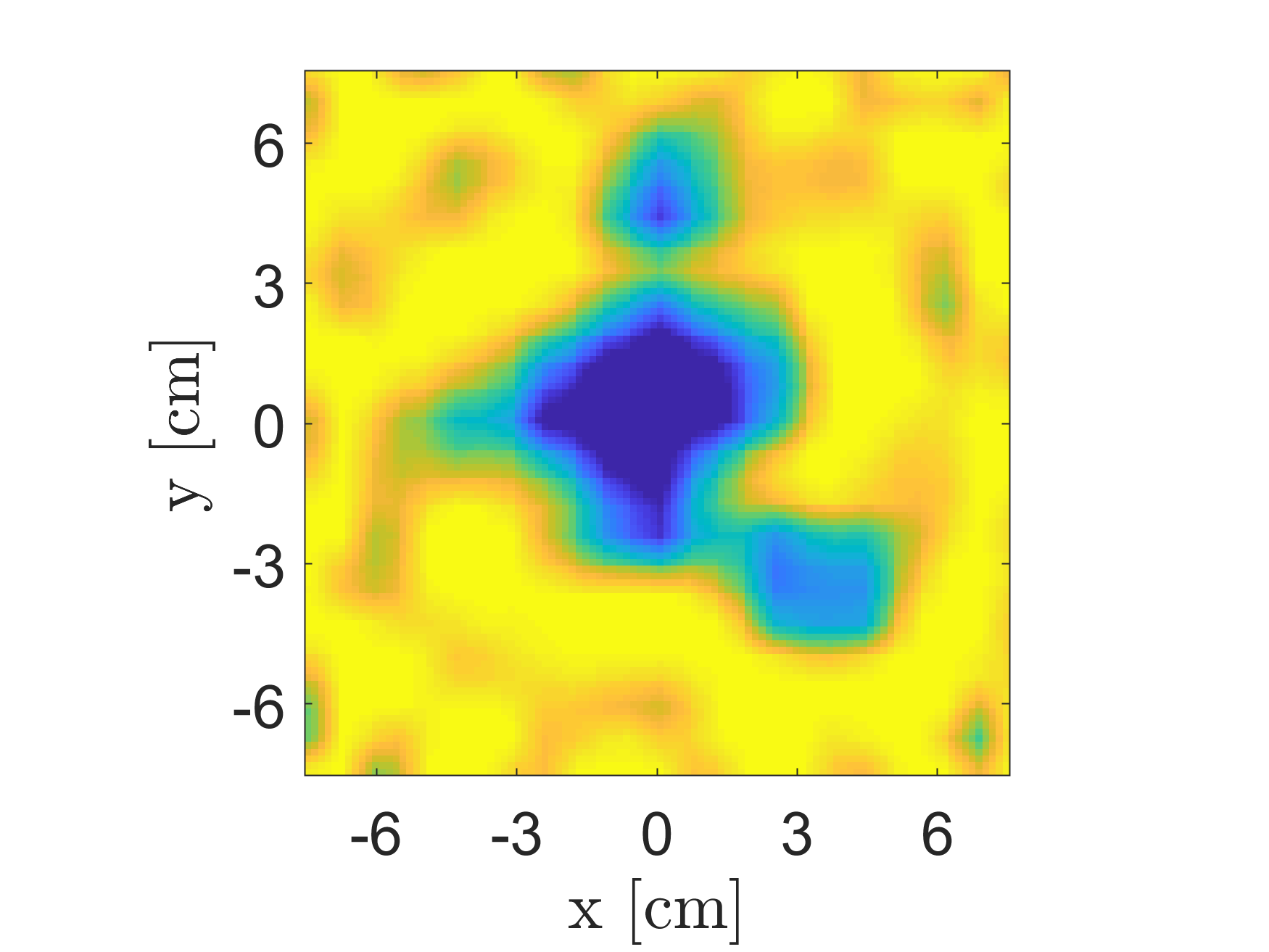}} & 
        {\includegraphics[draft=false, trim=2.6cm 1cm 2cm 0cm, clip, height=3.7cm]{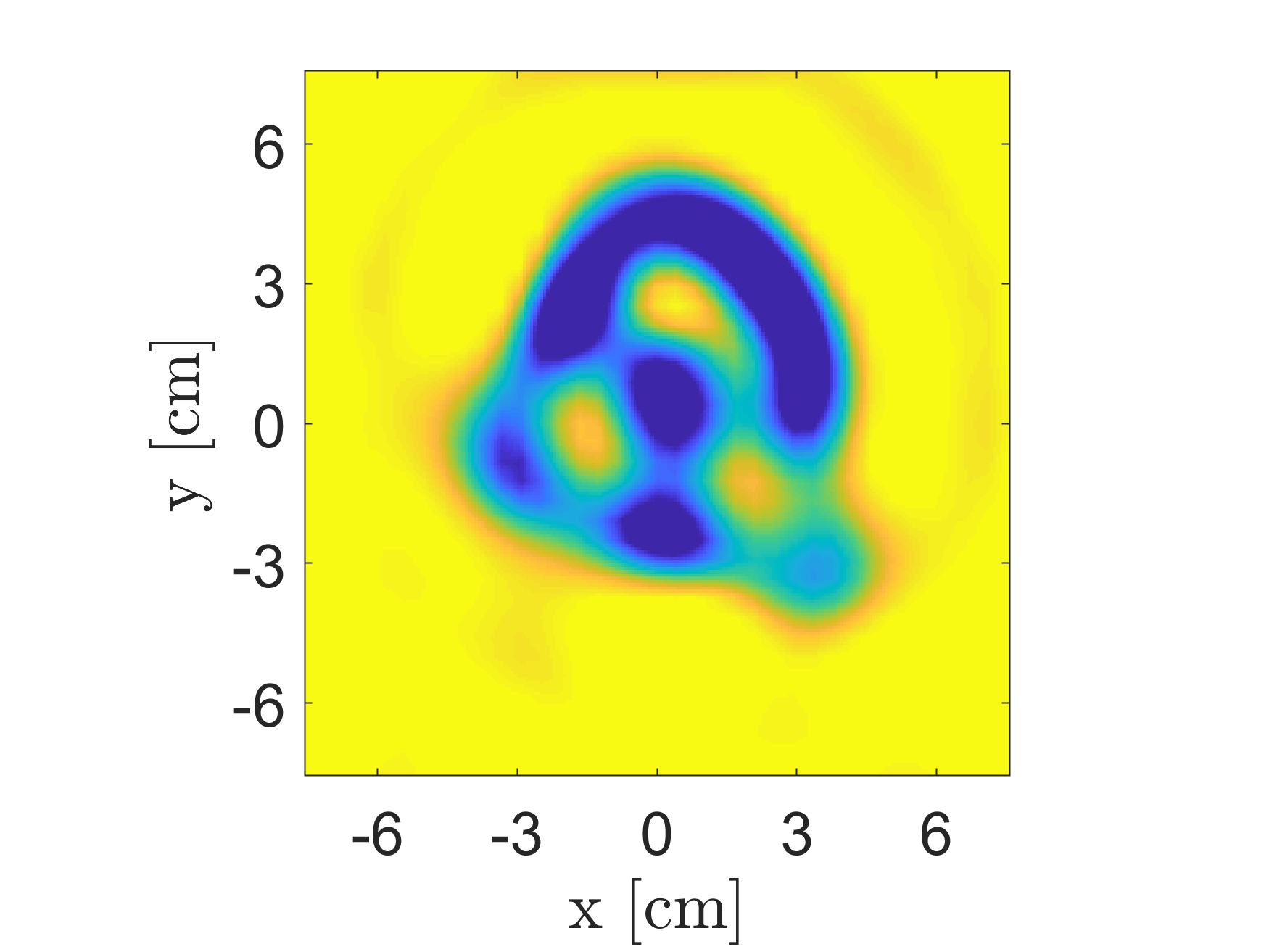}} & 
        {\includegraphics[draft=false, trim=2.6cm 1cm 2cm 0cm, clip, height=3.7cm]{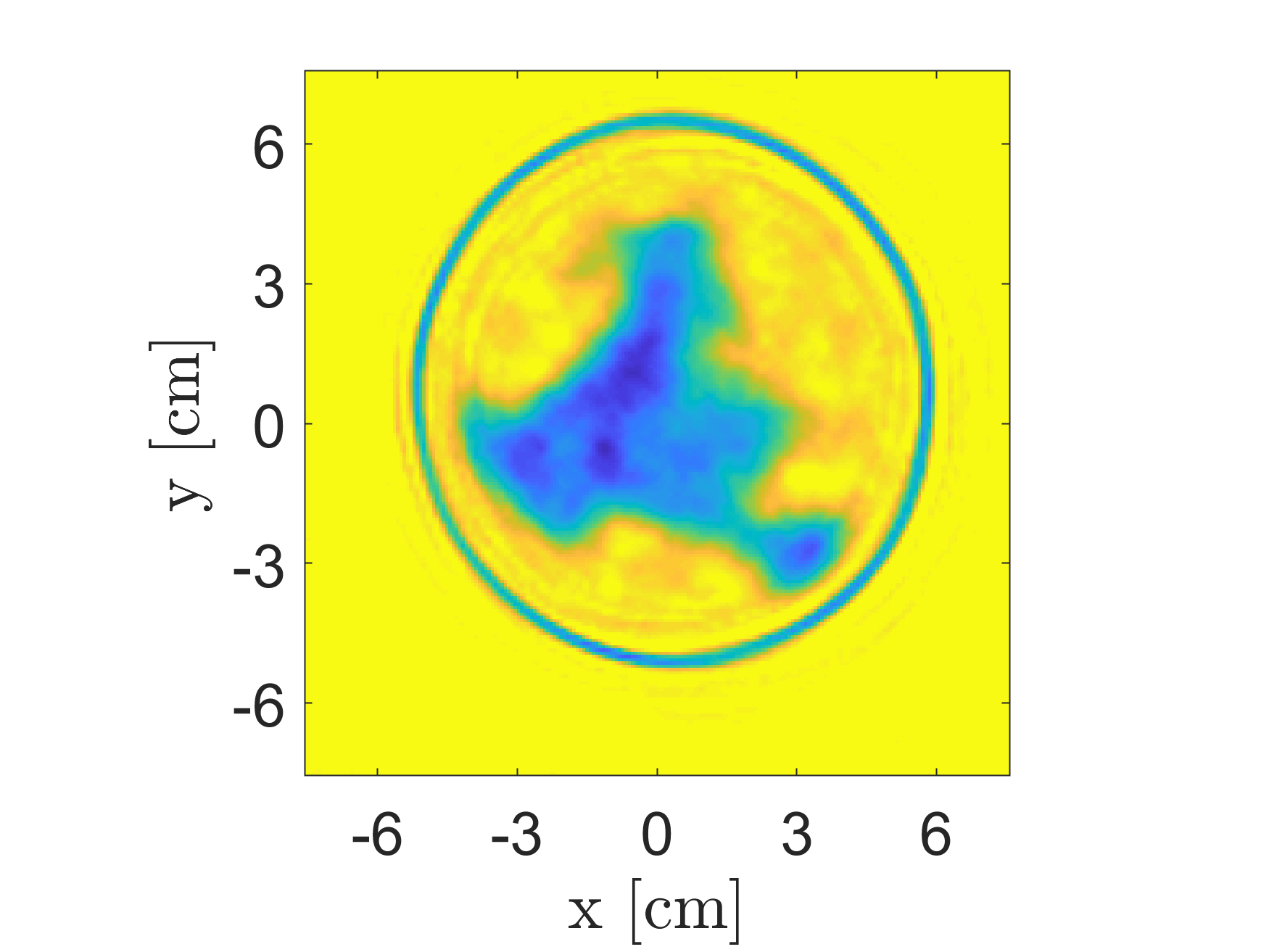}} &
        {\includegraphics[draft=false, trim=2cm 1cm 1.8cm 0cm, clip, height=3.7cm]{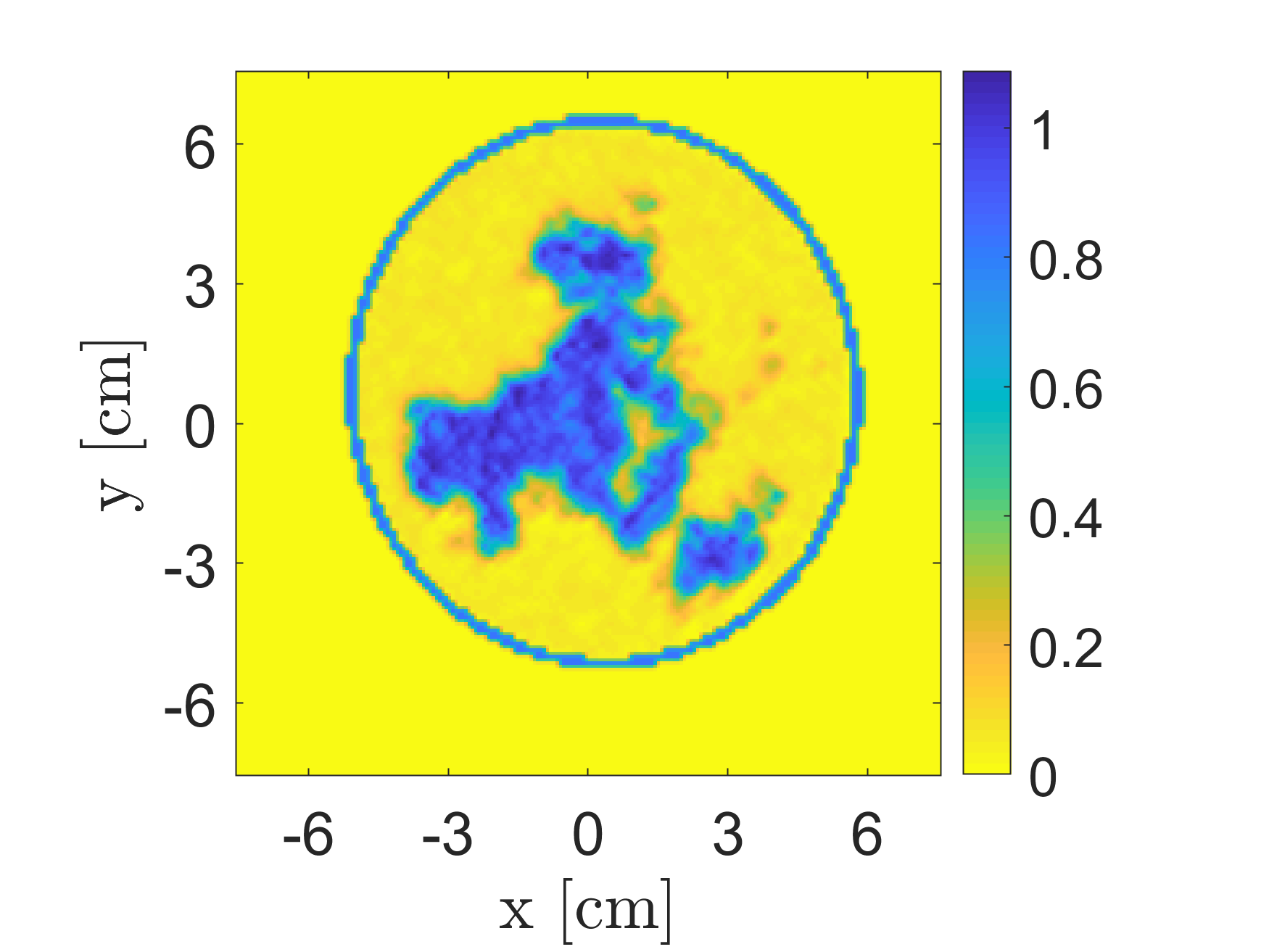}}  \\
 (e) DBIM & (f) CSI & (g) ANN & (h) Reference \\
        {\includegraphics[draft=false, trim=2.6cm 1cm 2cm 0cm, clip, height=3.7cm]{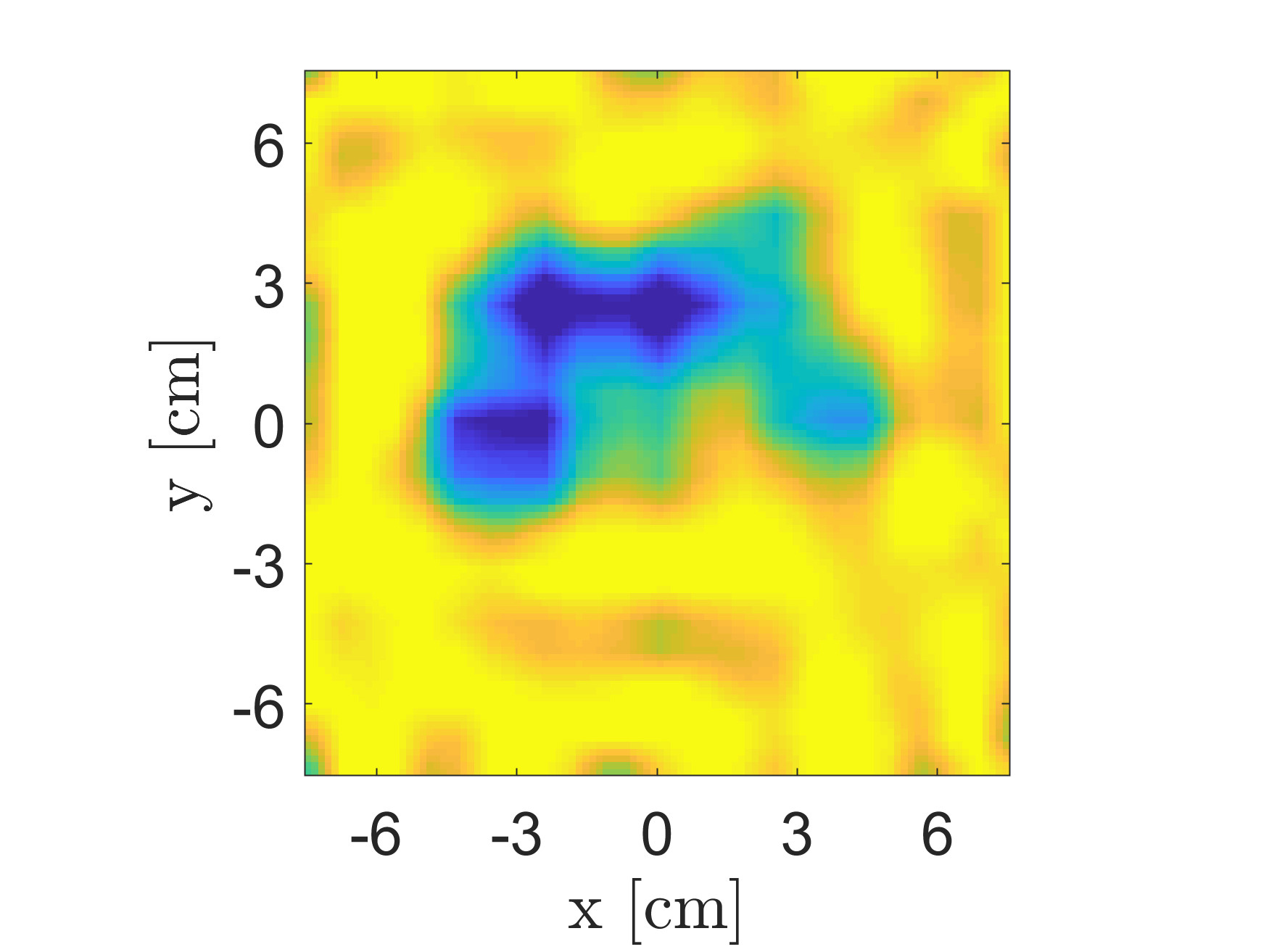}} & 
        {\includegraphics[draft=false, trim=2.6cm 1cm 2cm 0cm, clip, height=3.7cm]{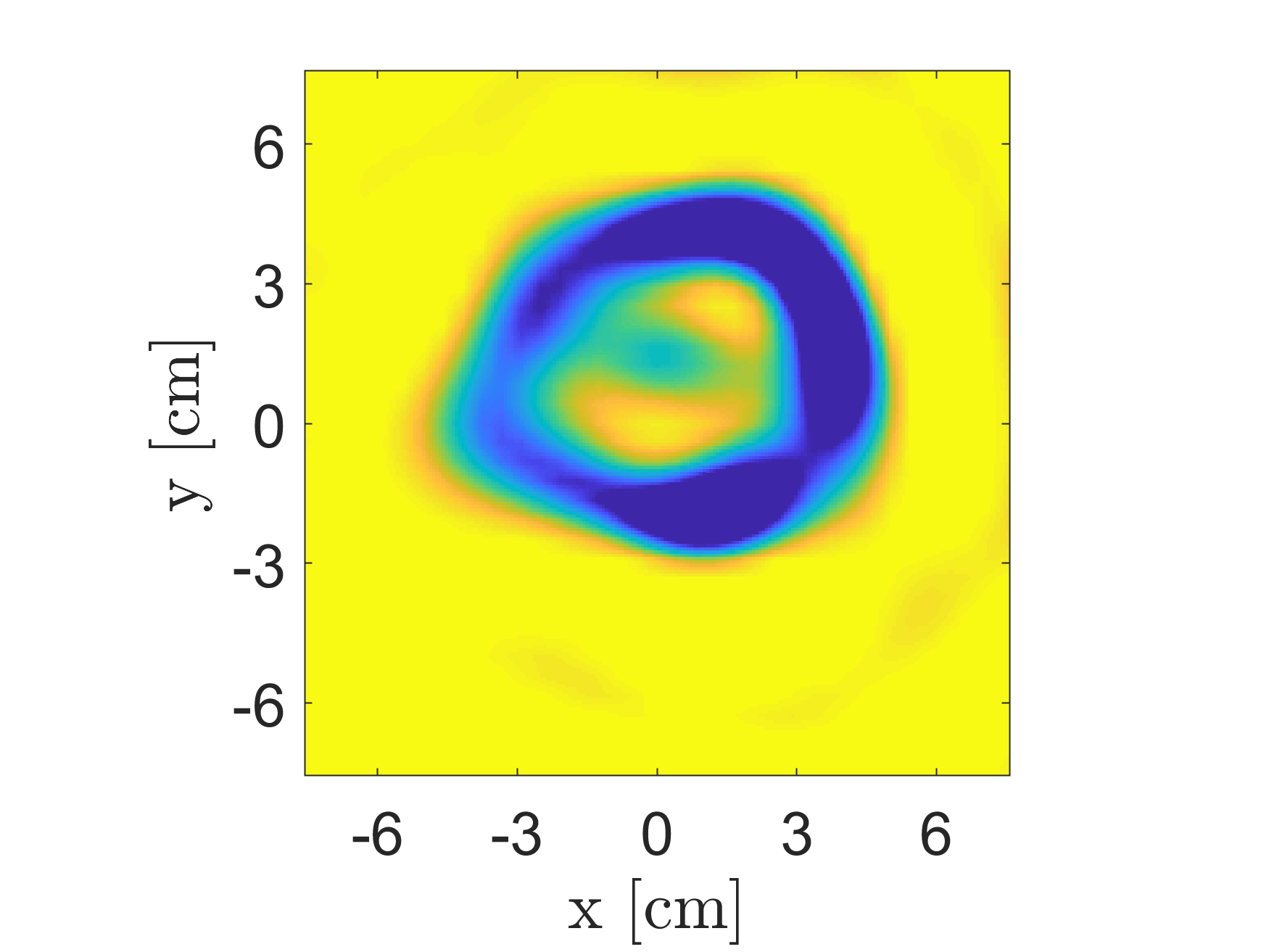}} & 
        {\includegraphics[draft=false, trim=2.6cm 1cm 2cm 0cm, clip, height=3.7cm]{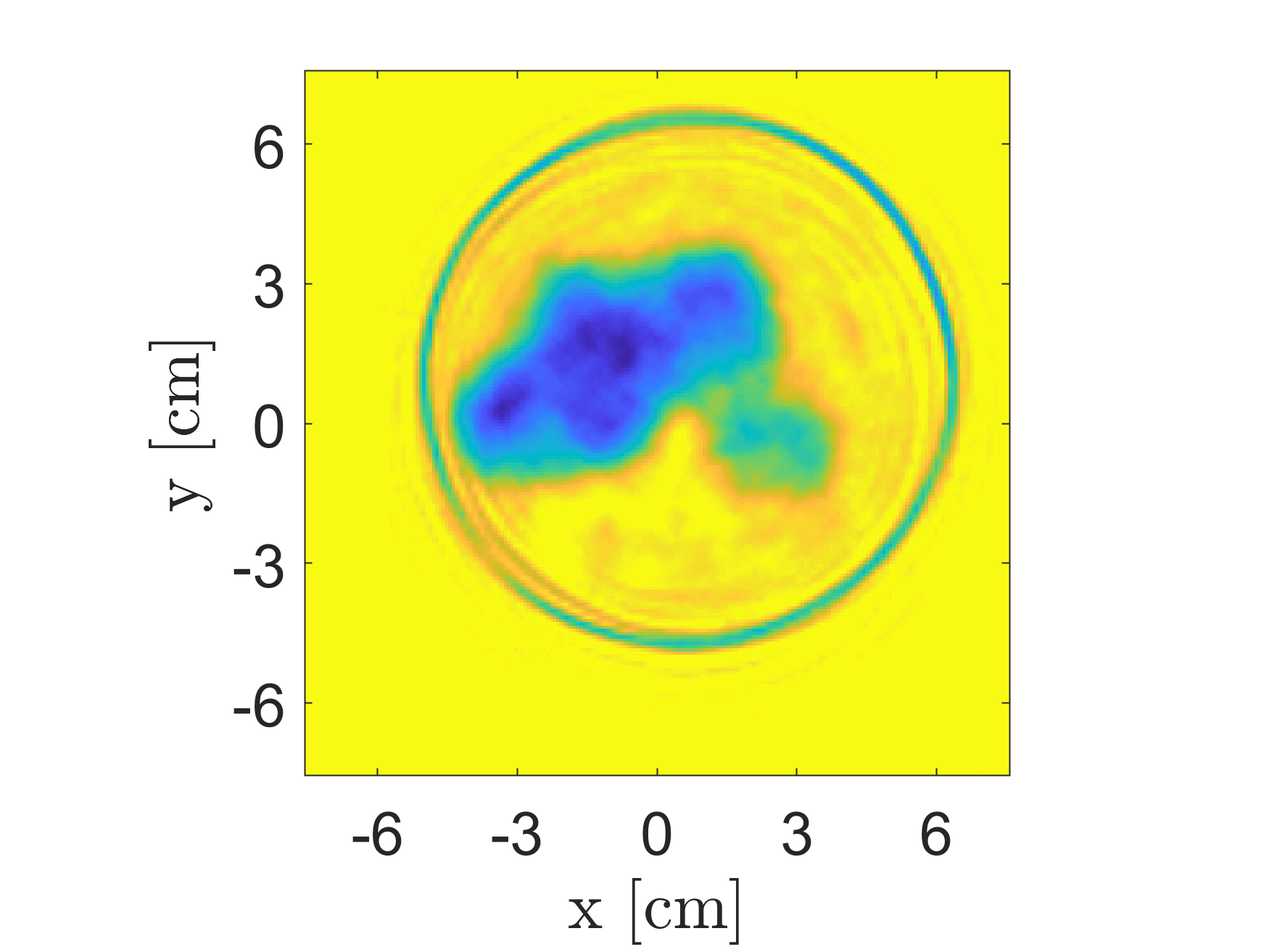}} &
        {\includegraphics[draft=false, trim=2cm 1cm 1.8cm 0cm, clip, height=3.7cm]{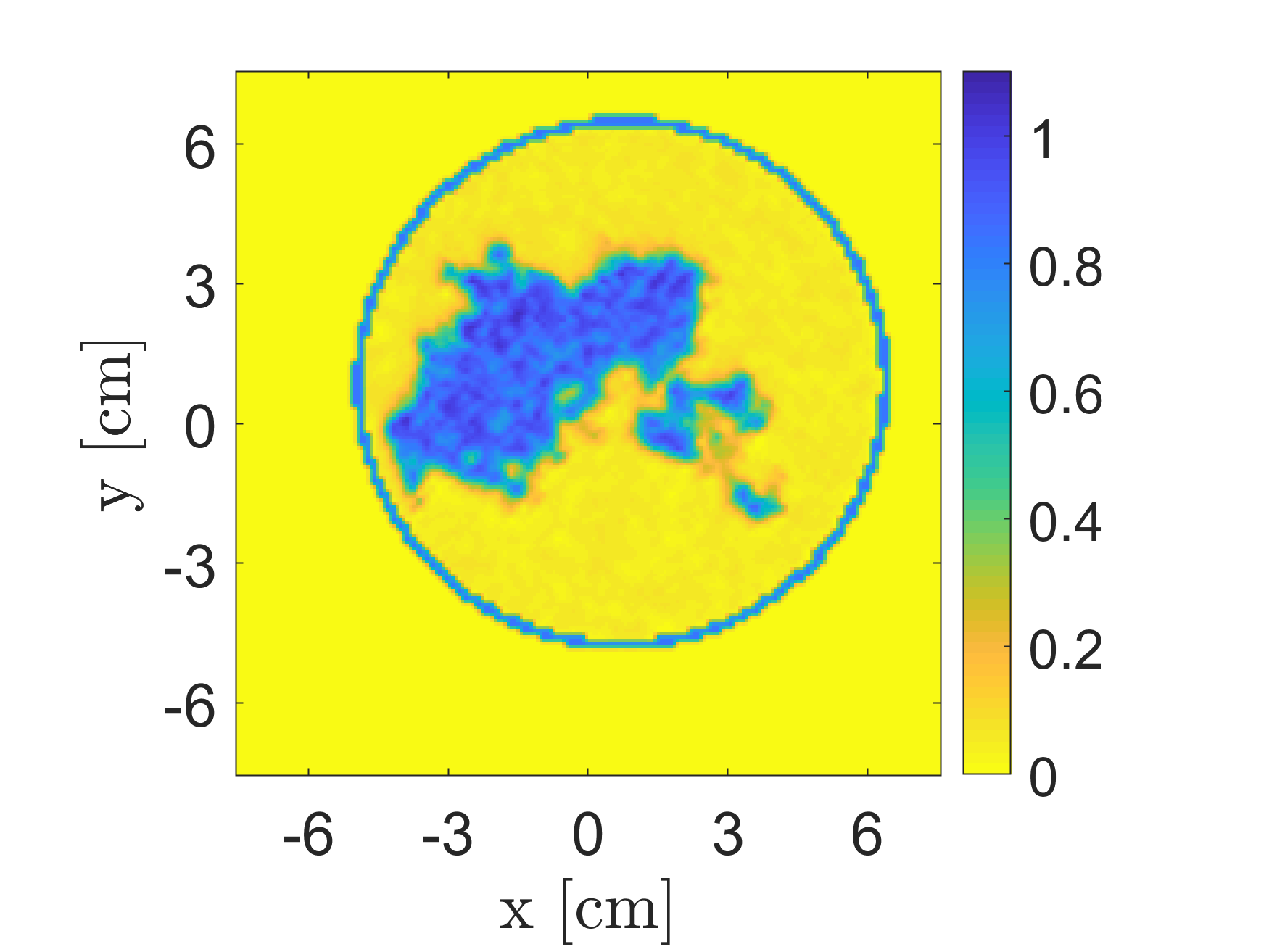}}  \\
(i) DBIM & (j) CSI & (k) ANN & (l) Reference \\
   \end{tabular}}
   \caption{Comparison between recoveries related to three different breast profiles. Retrieved conductivity maps (measure unit: S/m) for DBIM (a), (e), (i), CSI (b), (f), (j), and ANN-based approach (c), (g), (k). Reference conductivity maps are reported in (d), (h), (l). Both the x and y axes are in centimetre units.}  
   \label{fig:net_comparisons_im}
\end{figure*}

\subsection{Neural Network Design}
\label{sec:methodology:ANN}
%
Differently from other approaches presented in literature, 
in this manuscript a fully-connected architecture is considered 
instead of the commonly adopted CNNs.

The key concept of an EIS inversion consists in the fact that a direct mapping between two different spaces, i.e. the scattered field and the dielectric contrast, is established. Most of the work proposed in the scientific literature focuses on the use on CNNs architecture \cite{xu2020deep, li2018deepnis, wei2018deep}, but these architectures are very useful typically when the input information flow is local, i.e. each output value is related to an input subset, such as denoising or despeckling applications \cite{Zhang_2017, 9261137}. Conversely, in the case of EIS problems, there is not a direct link between the data and unknowns spaces, therefore a preliminary extra-mapping is usually required for reaching good accuracy and optimal inversion performance. Practically, an initial raw inversion is mostly performed to move from the data to the unknowns space, and then a CNN architecture is usually applied which represents the second part of a two-step inversion procedure \cite{wei2018deep}.
Thus, for imaging purposes, the fully connected ANNs feature of having global links between all nodes of consecutive layers seems the key aspect to us.
In the network architectures considered in the following, an adaptive moment estimation method (ADAM) was employed to minimize the cost function with an initial learning rate of $5\cdot 10^{-5}$ and 30 epochs per each training phase. 

The optimal layout of the network in terms of number of hidden layers and neurons has been identified as first step. The analysis is reported in Section \ref{sec:result}.



\section{Results}
\label{sec:result}
%
%
In this section a numerical analysis to design properly the network architecture, i.e. the number of nodes and layers, will be addressed. Moreover, the methodology is tested in case of different scenarioes. The recovery performance is evaluated by comparing the proposed ANN-based reconstructions with those obtained via conventional nonlinear approaches. 
\begin{figure*}[t]
   \centerline{ \begin{tabular}{cccc}
        {\includegraphics[draft=false, trim=2.6cm 1cm 1.6cm 0cm, clip, height=3.7cm]{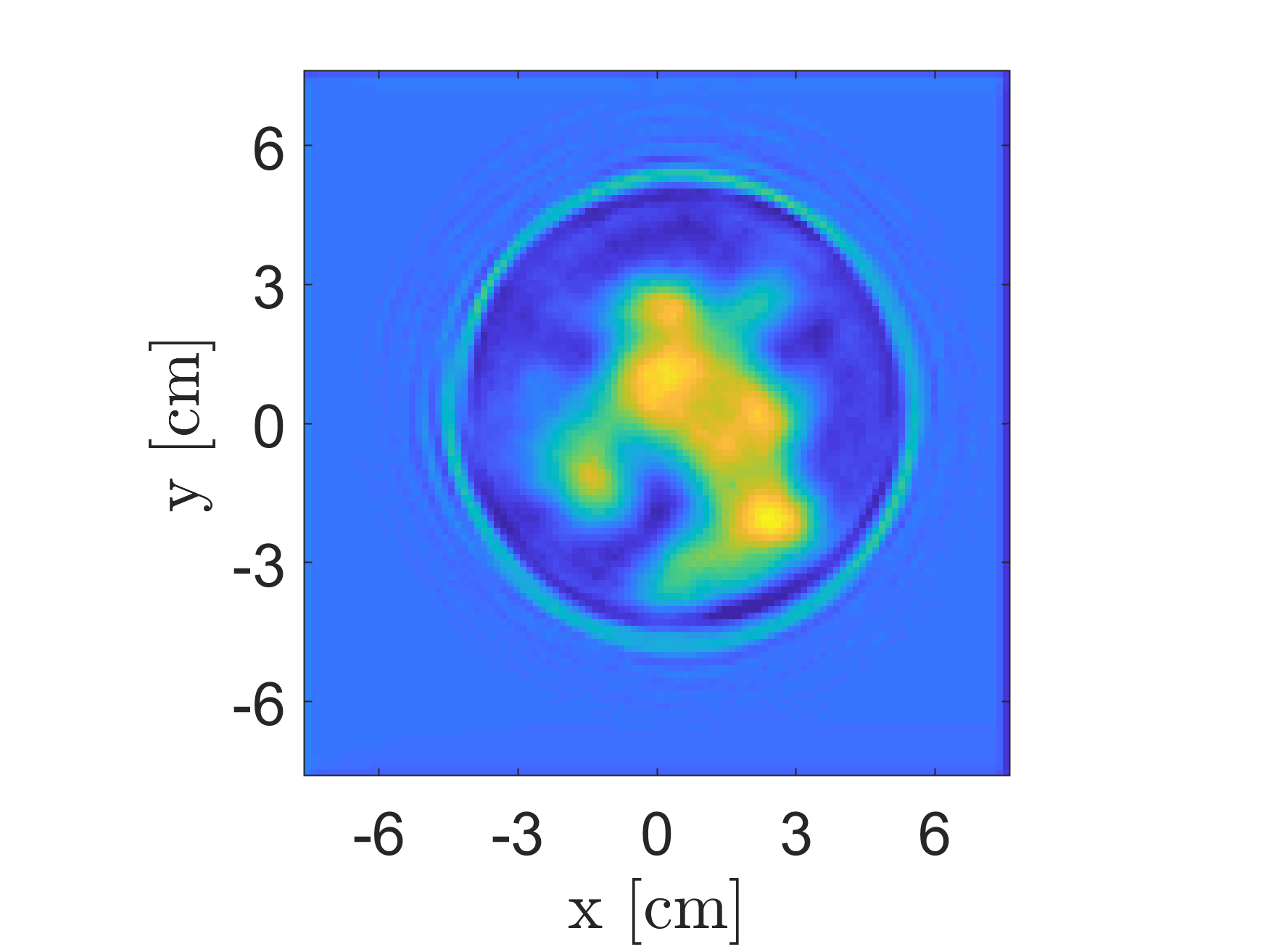}} & 
        {\includegraphics[draft=false, trim=2cm 1cm 1.6cm 0cm, clip, height=3.7cm]{Results/Ref2_re.png}} & 
        {\includegraphics[draft=false, trim=2.6cm 1cm 1.6cm 0cm, clip, height=3.7cm]{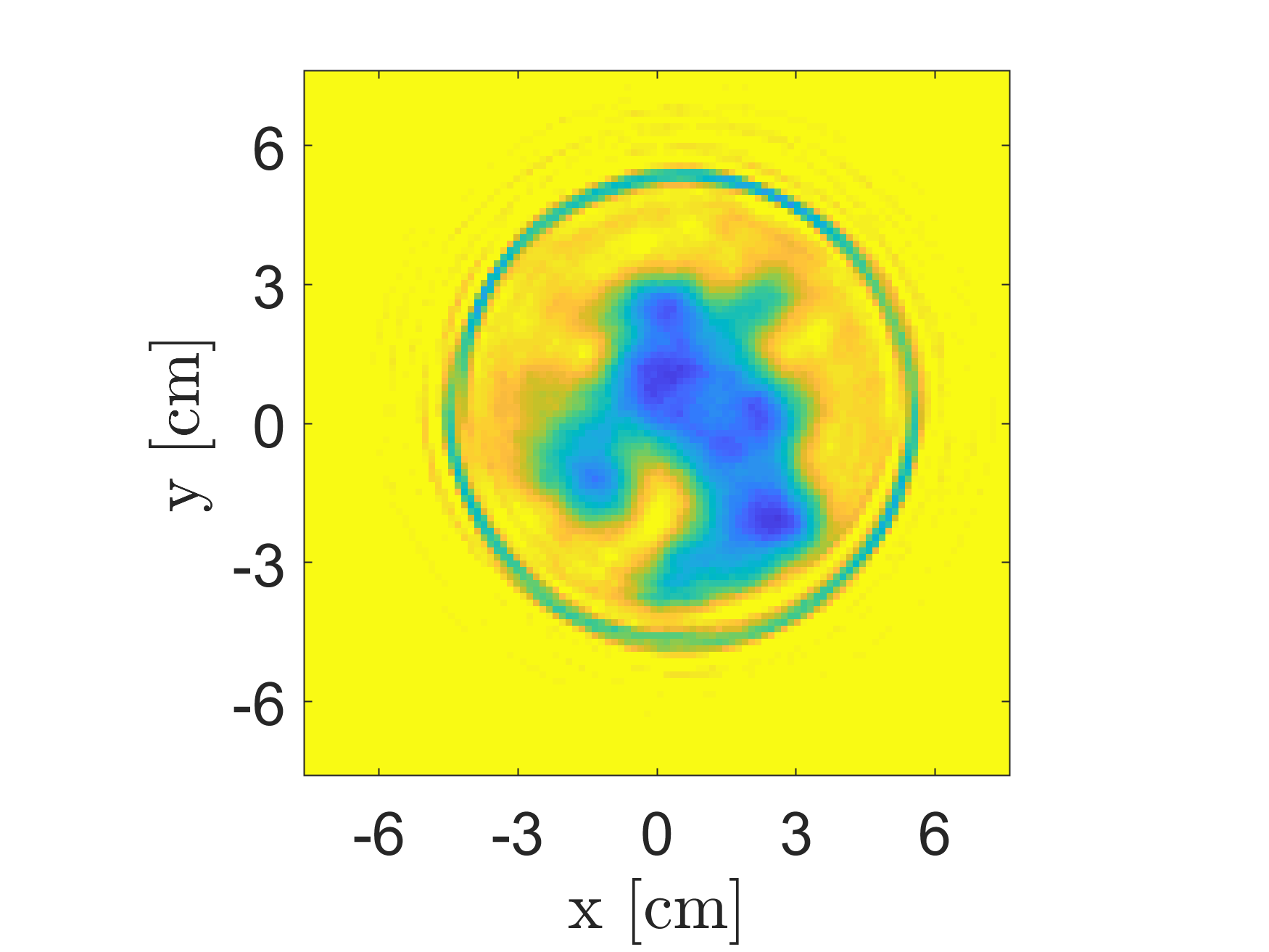}} &
        {\includegraphics[draft=false, trim=2cm 1cm 1.6cm 0cm, clip, height=3.7cm]{Results/Ref2_sig.png}}  \\
 (a) SNR = 5 dB & (b) Reference & (c) SNR = 5 dB & (d) Reference\\
         {\includegraphics[draft=false, trim=2.6cm 1cm 1.6cm 0cm, clip, height=3.7cm]{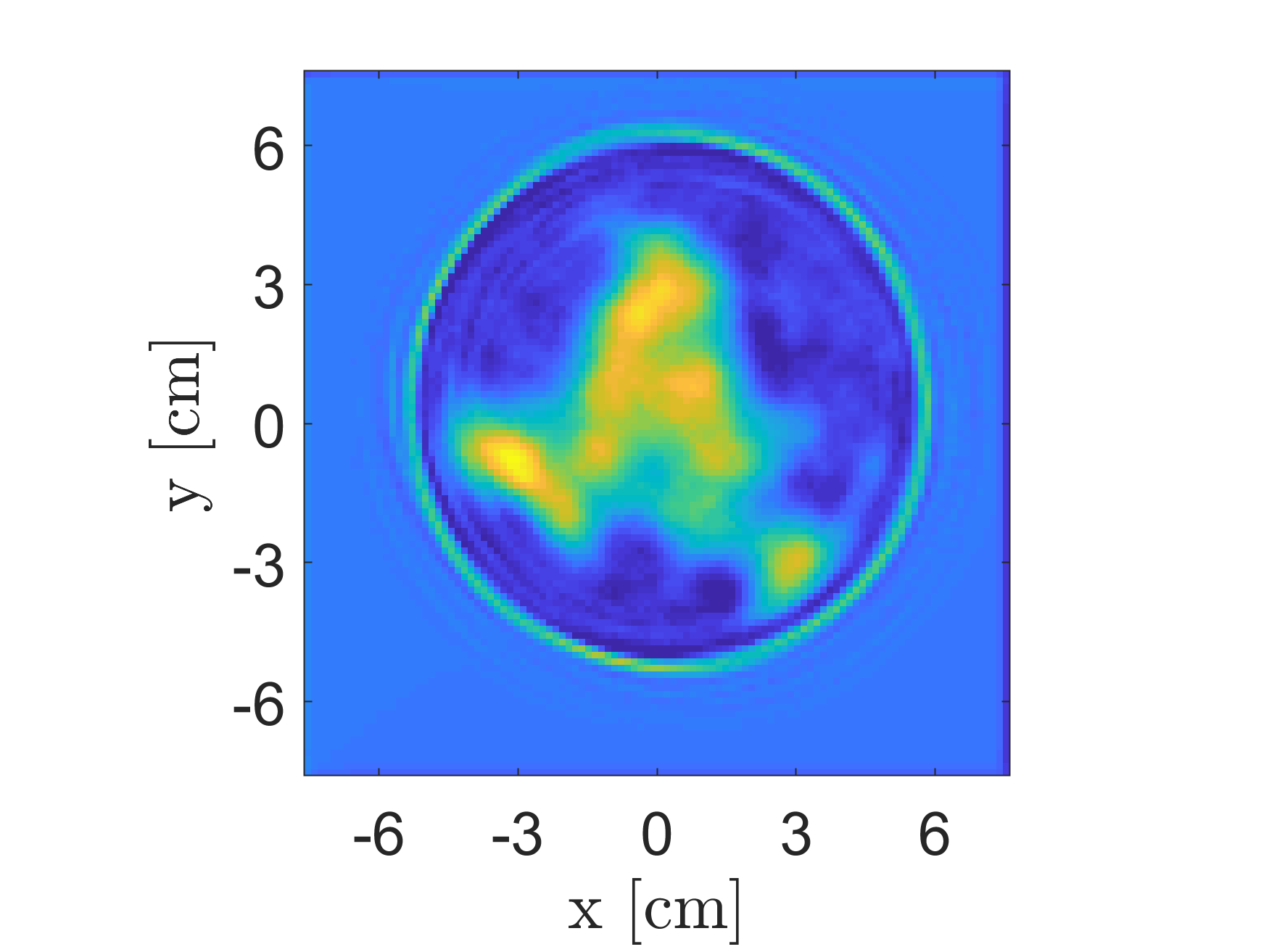}} & 
        {\includegraphics[draft=false, trim=2cm 1cm 1.6cm 0cm, clip, height=3.7cm]{Results/Ref5_re.png}} & 
        {\includegraphics[draft=false, trim=2.6cm 1cm 1.6cm 0cm, clip, height=3.7cm]{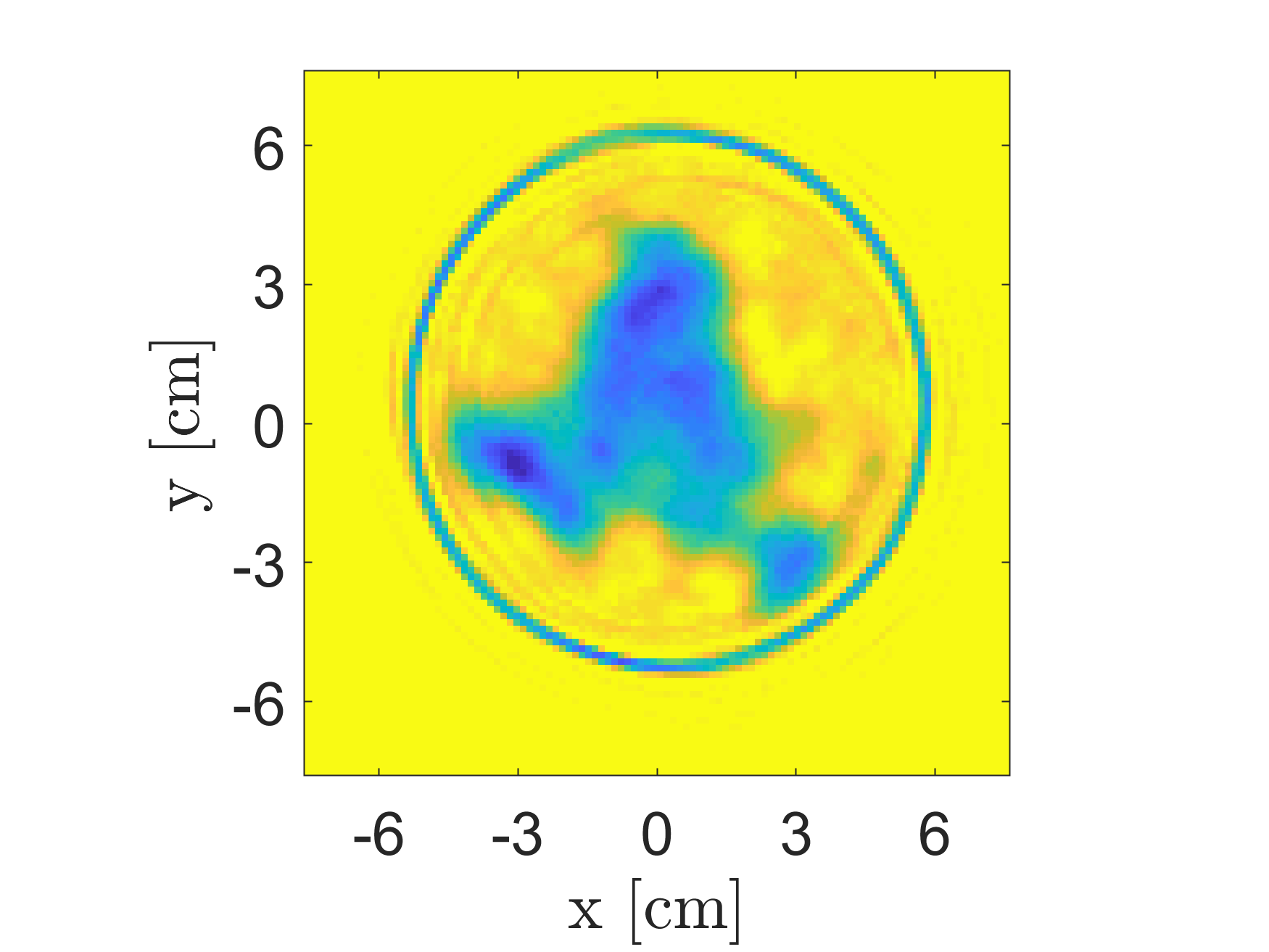}}  &
        {\includegraphics[draft=false, trim=2cm 1cm 1.6cm 0cm, clip, height=3.7cm]{Results/Ref5_sig.png}}  \\
 (e) SNR = 5 dB & (f) Reference & (g) SNR = 5 dB & (h) Reference \\
         {\includegraphics[draft=false, trim=2.6cm 1cm 1.6cm 0cm, clip, height=3.7cm]{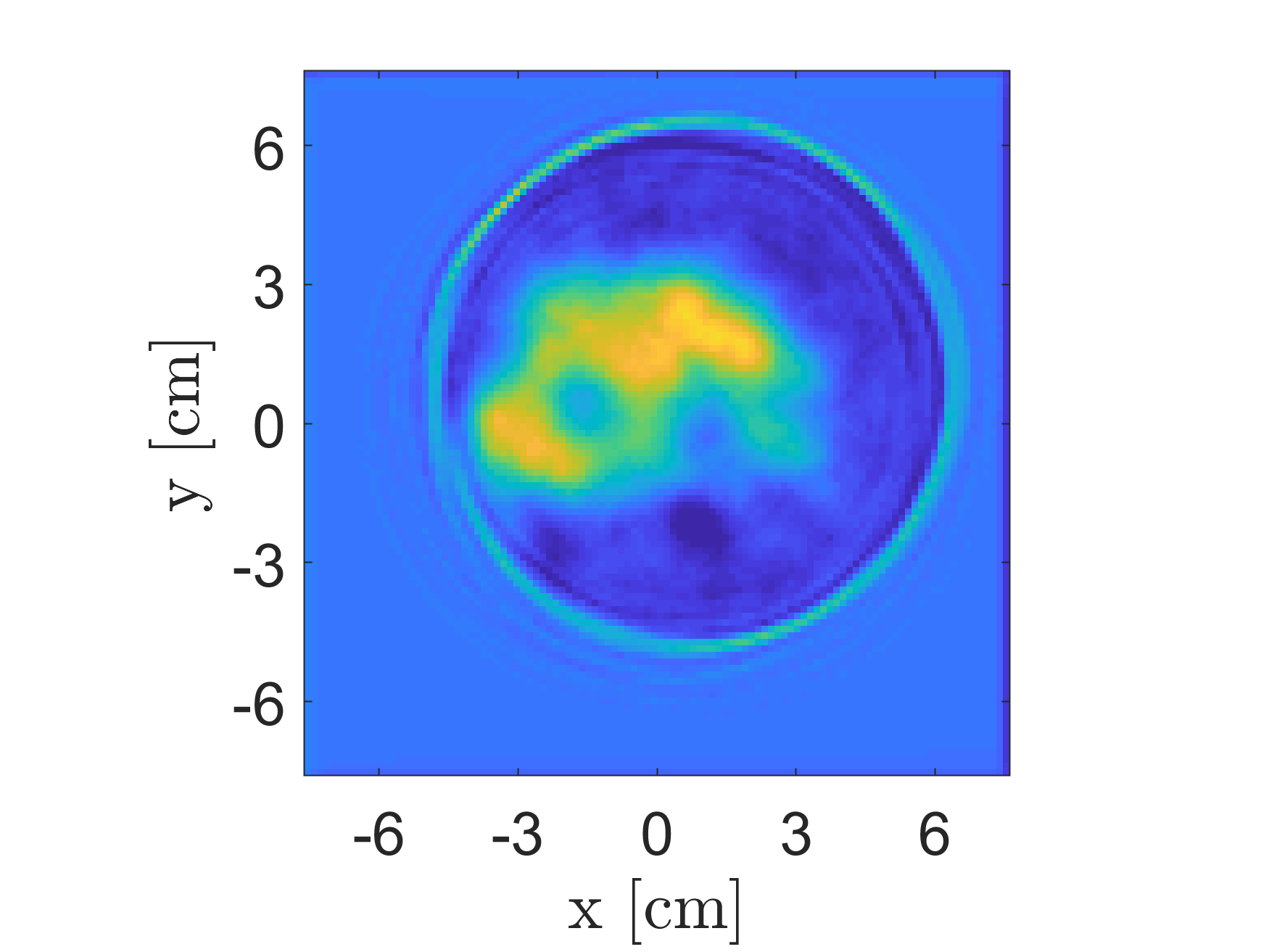}} & 
        {\includegraphics[draft=false, trim=2cm 1cm 1.6cm 0cm, clip, height=3.7cm]{Results/Ref7_re.png}} & 
        {\includegraphics[draft=false, trim=2.6cm 1cm 1.6cm 0cm, clip, height=3.7cm]{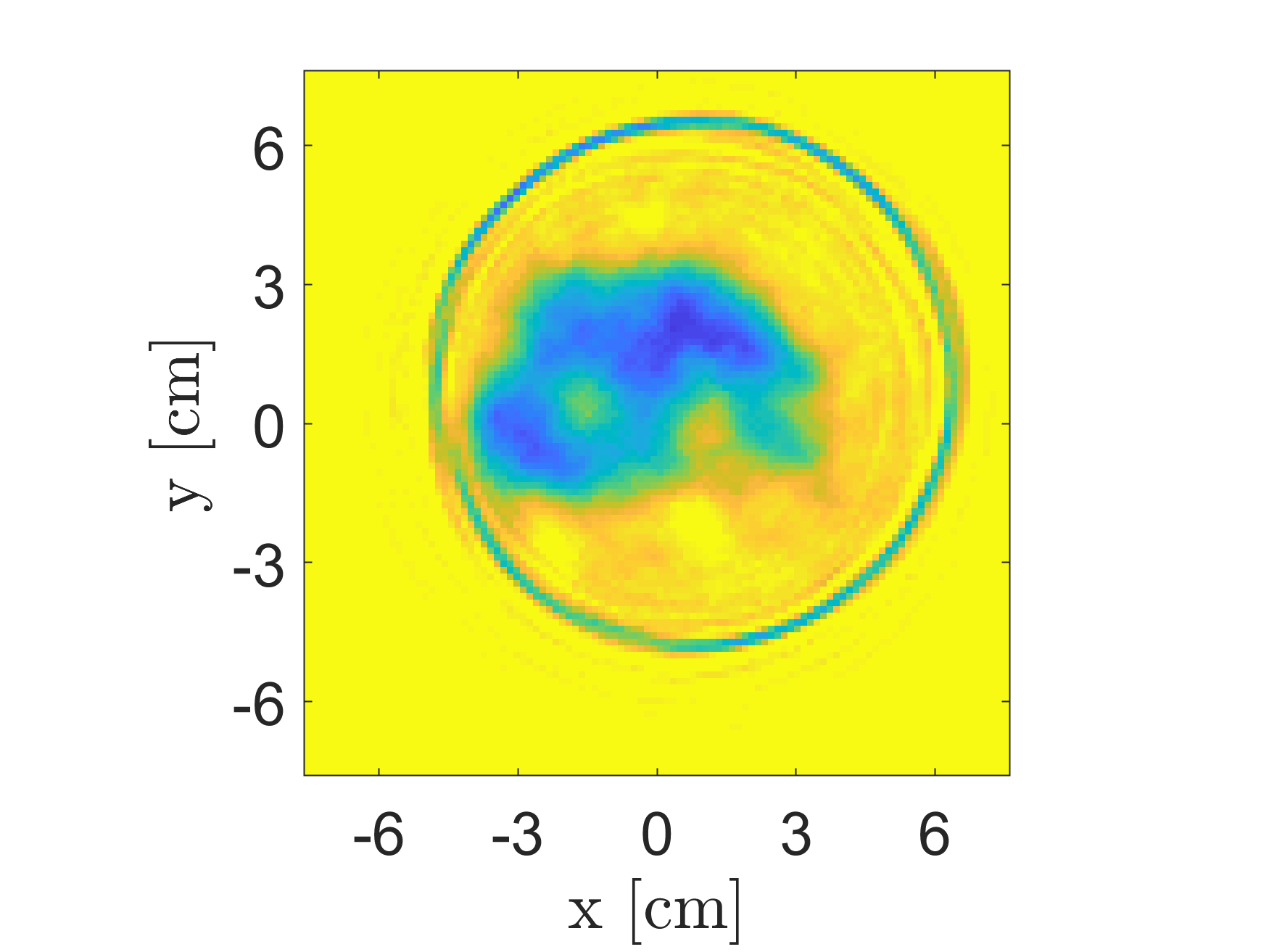}} &
        {\includegraphics[draft=false, trim=2cm 1cm 1.6cm 0cm, clip, height=3.7cm]{Results/Ref7_sig.png}}  \\
 (i) SNR = 5 dB & (j) Reference & (k) SNR = 5 dB & (l) Reference \\
    \end{tabular}}
   \caption{Performance assessment of the proposed ANN-based approach for an SNR = 5 dB. Retrieved permittivity (a, e, i) and conductivity maps (c, g, k), and corresponding reference profiles (b, f, j, d, h, l). Both the x and y axes are in centimetre units.}  
   \label{fig:perf_ass_VS_noise}
\end{figure*}

\subsection{Network Training}
In order to perform a numerical assessment, a quantitative, fair metric to determine the quality of the recoveries is required.
Conventionally in the image processing community parameters such as the normalised mean square error (NMSE) and the structural similarity index measure (SSIM) are considered. Unfortunately, these kinds of metrics proved to be not much effective for EIS problems performance assessment \cite{nikolova2011microwave, benny2020overview, aldhaeebi2020review}. 
Thus, we decided to evaluate the capability of the approach in terms of details retrieving by evaluating the bandwidth of the reconstructed profile. This approach can be considered as the evaluation of the filtering properties of combination of the inverse radiating operator and the imaging algorithm.

For every image of the testing set, the 2D power spectrum in polar coordinates $(\nu,\theta)$ was evaluated and averaged with respect to all the images spectra, obtaining a mean-squared spectrum $\mathcal{T}(\nu,\theta)$.
By averaging along the angle coordinate, a one-dimensional spectrum $\mathcal{S}(\nu)$ is obtained, i.e.:

\begin{align}
\label{eq:2D_MSS}
    &\mathcal{T}(\nu,\theta)=\frac{1}{N}\sum_{n=1}^{N}{|\mathscr{F}(\Tilde{x}_n)}|^2\,, \\
\label{eq:1D_MSS}
    &\mathcal{S}(\nu)=\frac{1}{2\pi}\int_{-\pi}^{\pi} \mathcal{T}(\nu,\theta)d\theta\,,
\end{align}
in which $N=6000$ is the number of testing profiles, $\mathscr{F}$ is the Fourier transform operator, $\Tilde{x}_n$ is the retrieved estimate of the breast profile obtained via ANN inversion. 


The motivation supporting this performance indicator lies in the fact that the closer the recovery is to the true profile, the closer their spectral information will be, providing an indirect measure of the filtering properties of the inverse radiating operator and of its impact at each frequency. This procedure clearly yields a spectral comparison rather than a direct similarity measure.

In order to find a trade-off between computational complexity and accuracy in the recovery, a performance assessment on different network topologies was carried out. As first analysis, the impact of the number of nodes has been evaluated in case of
a three-layer fully-connected network. As expected, Fig. \ref{fig:mean_spect} shows that the higher the number of nodes, the better the quality of the recovery since the averaged spectrum $\mathcal{S}(\nu)$ tends to the ideal behaviour of the true profile. Nevertheless, the higher is the number of nodes, the higher is the computational burden which yields to longer training time. Thus, the 2000-node case was selected since the improvement in averaged spectrum trend started to be negligible from this topology. This is clearly visible if a -3-dB threshold is considered, for which no relevant difference can be identified between the 2000-node and 5000-node cases up to a numerical frequency of 0.24 Hz. 

Complementary to the previous analysis, a similar study about the number of layers has been performed. To this aim, the averaged spectrum $\mathcal{S}(\nu)$ is reported while increasing the number of hidden layers from one to five and adopting 2000 nodes per each layer.

The related results are shown in Fig. \ref{fig:mean_spect_layers}.
Also in this case no clear difference is visible by fixing a -3-dB threshold among the networks using more than three layers. Thus, in order to keep the computational complexity and processing time as low as possible, a three-layer architecture was selected (except the last regression layer not considered in these analyses).   
It is worth to note that we selected a number of breast profiles in the data set high enough for ensuring an effective training also in case of architectures with higher number of nodes/layers than the considered ones.

The computational time for the training of each architecture is a function of the number of involved layers and varies from 3 hours (1 layer) up to 8 hours (5 layers), while the inversion phase is almost real-time. These evaluations are performed on a Linux 64 bit workstation with an AMD Ryzen 3990X processor and an NVIDIA Quadro RTX 6000 graphics card.

\begin{figure*}[t]
   \centerline{ 
   \begin{tabular}{cccc}
        {\includegraphics[draft=false, trim=2.6cm 1cm 1.6cm 0cm, clip, height=3.7cm]{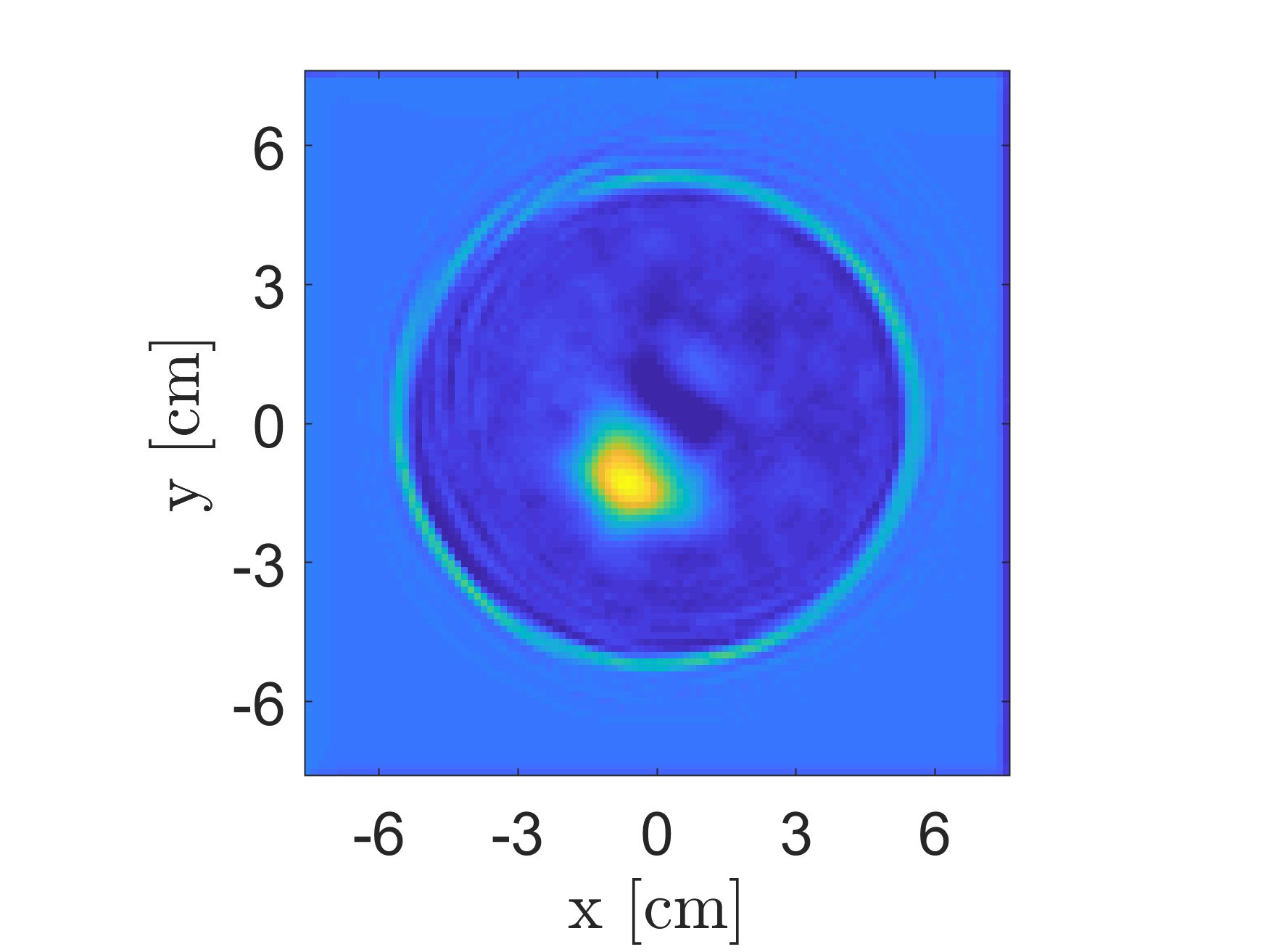}} & 
        {\includegraphics[draft=false, trim=2cm 1cm 1.6cm 0cm, clip, height=3.7cm]{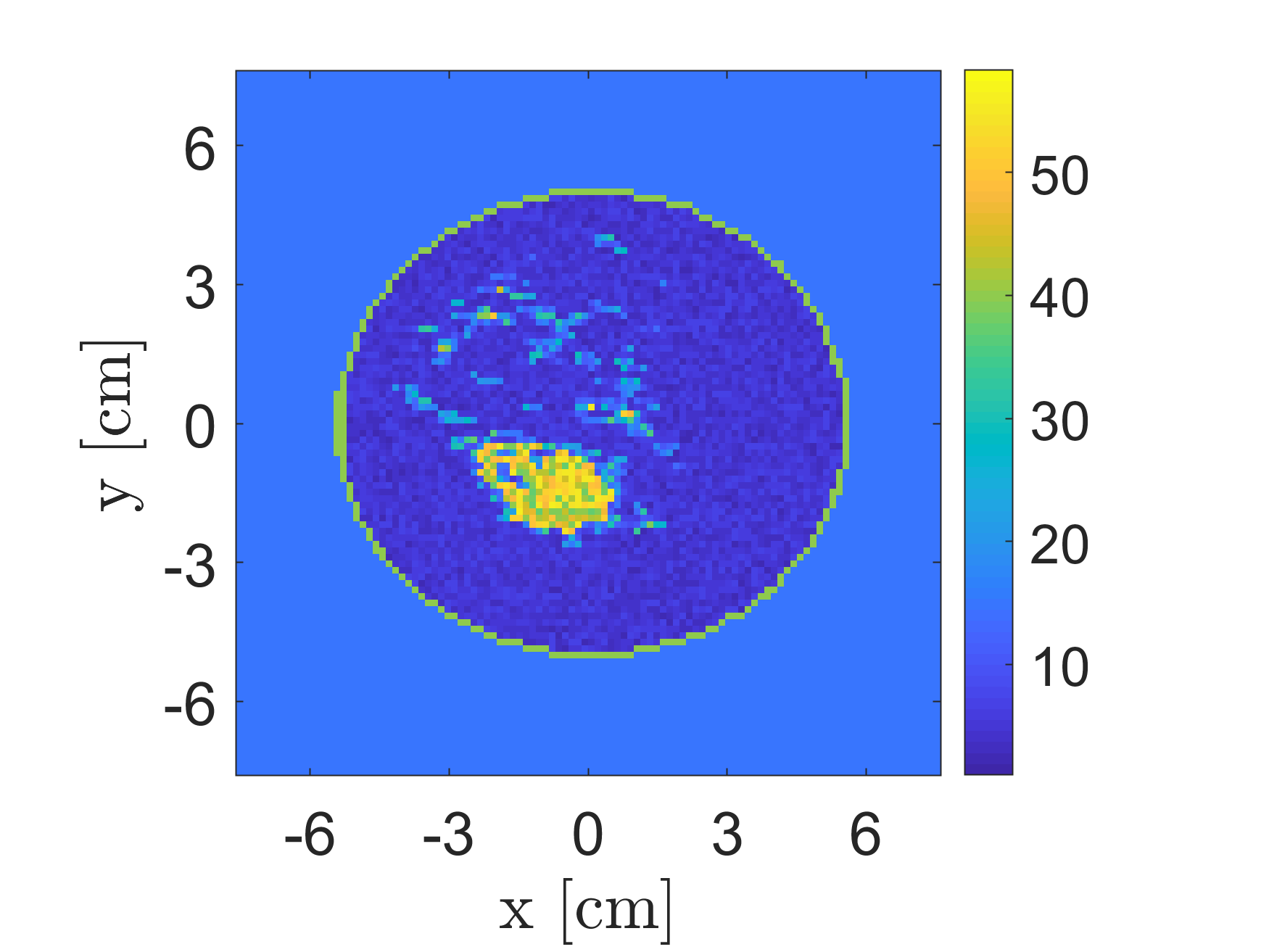}} &
        {\includegraphics[draft=false, trim=2.6cm 1cm 1.6cm 0cm, clip,  height=3.7cm]{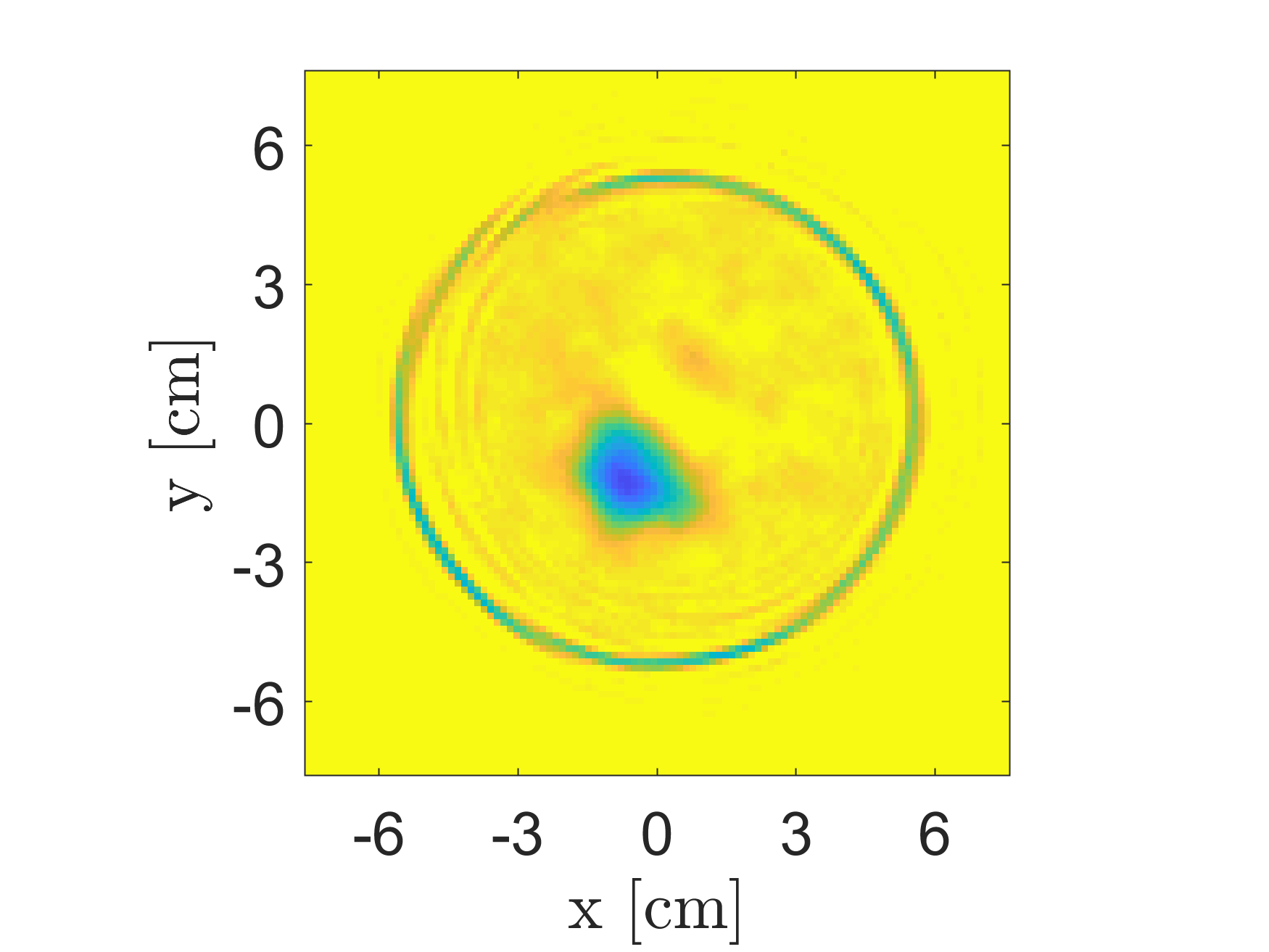}} &
        {\includegraphics[draft=false, trim=2cm 1cm 1.6cm 0cm, clip, height=3.7cm]{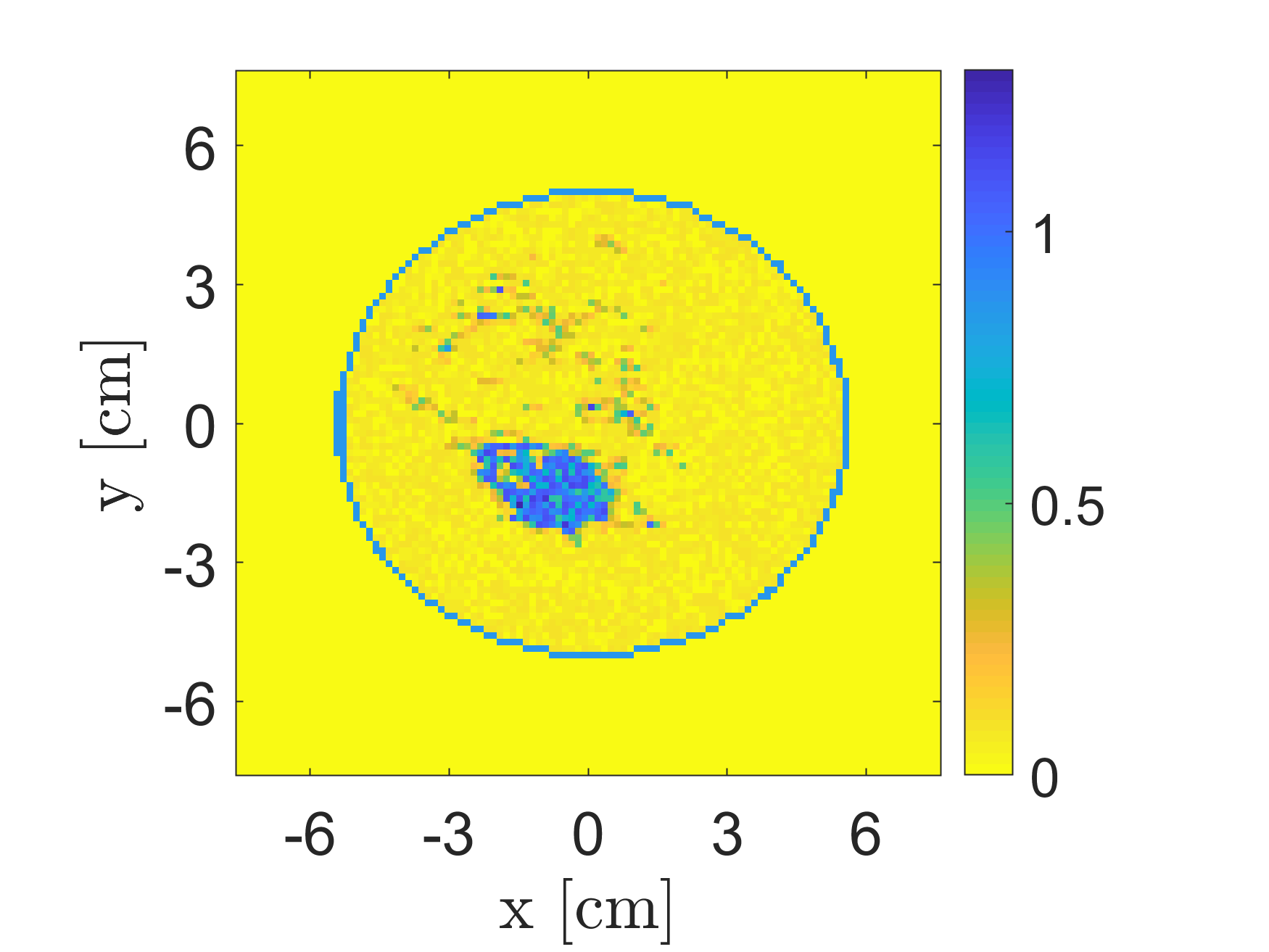}}  \\
 (a) & (b) & (c) & (d) \\   
        {\includegraphics[draft=false, trim=2.6cm 1cm 1.6cm 0cm, clip, height=3.7cm]{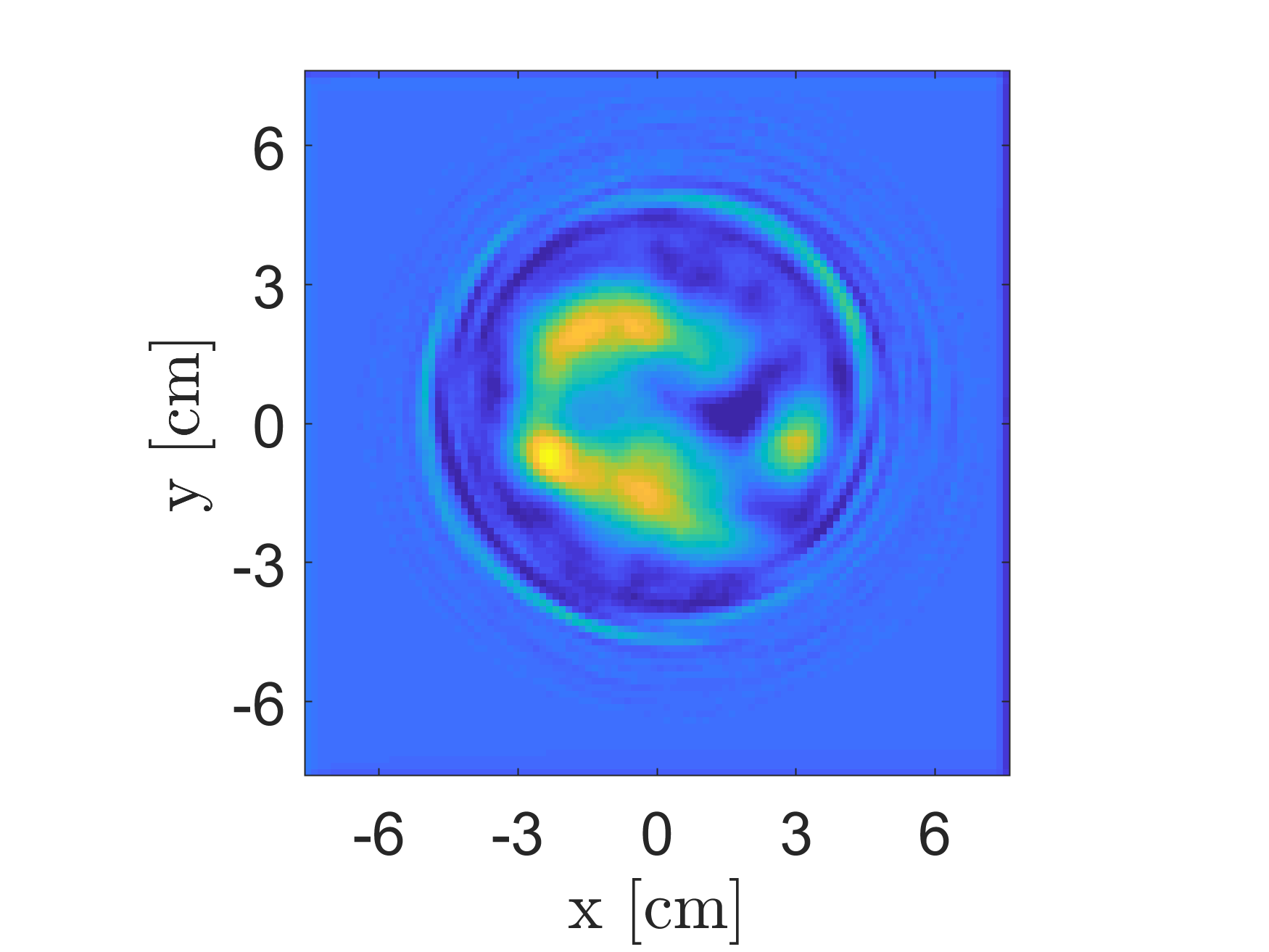}} & 
        {\includegraphics[draft=false, trim=2cm 1cm 1.6cm 0cm, clip, height=3.7cm]{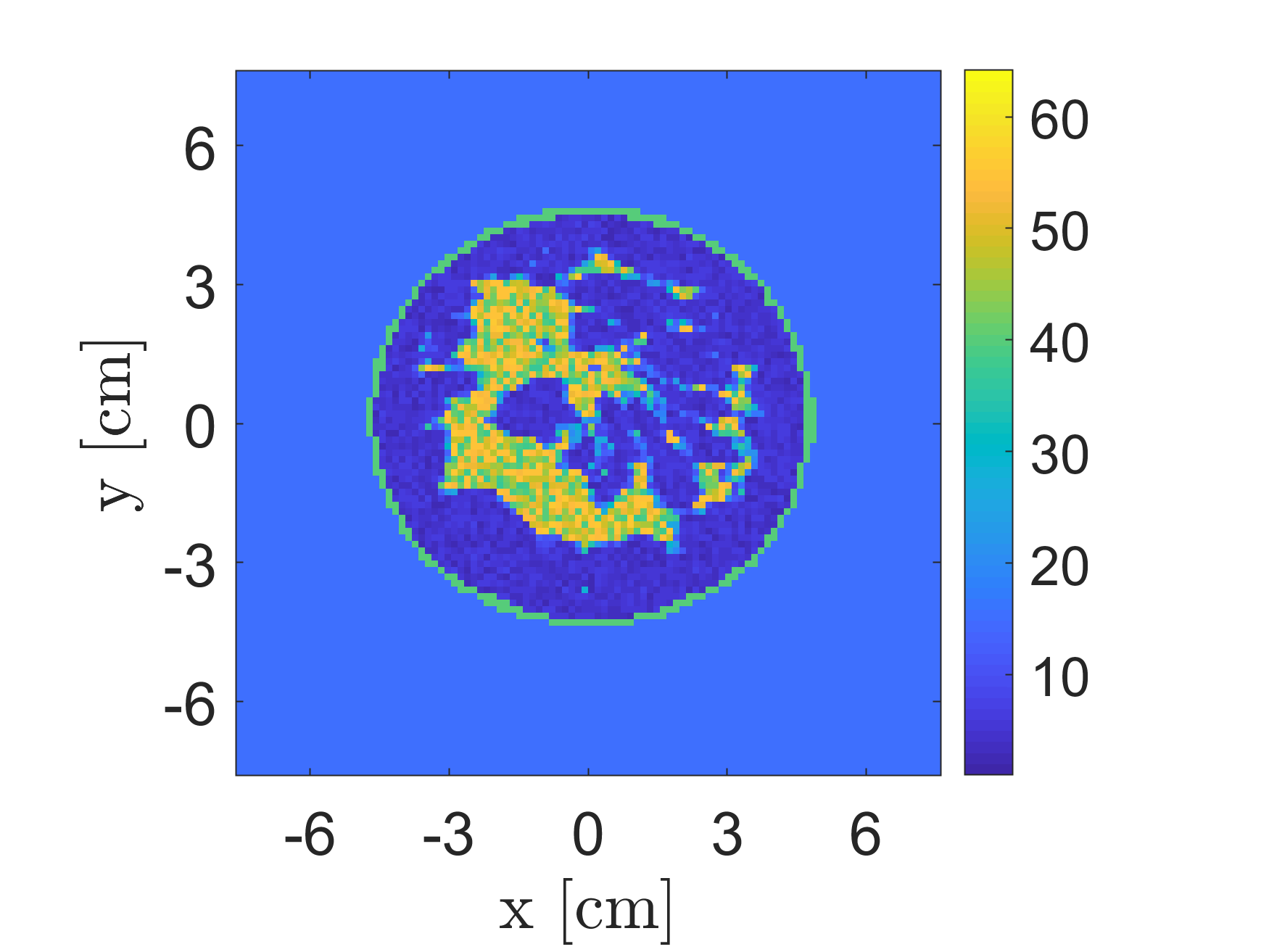}} &
        {\includegraphics[draft=false, trim=2.6cm 1cm 1.6cm 0cm, clip, height=3.7cm]{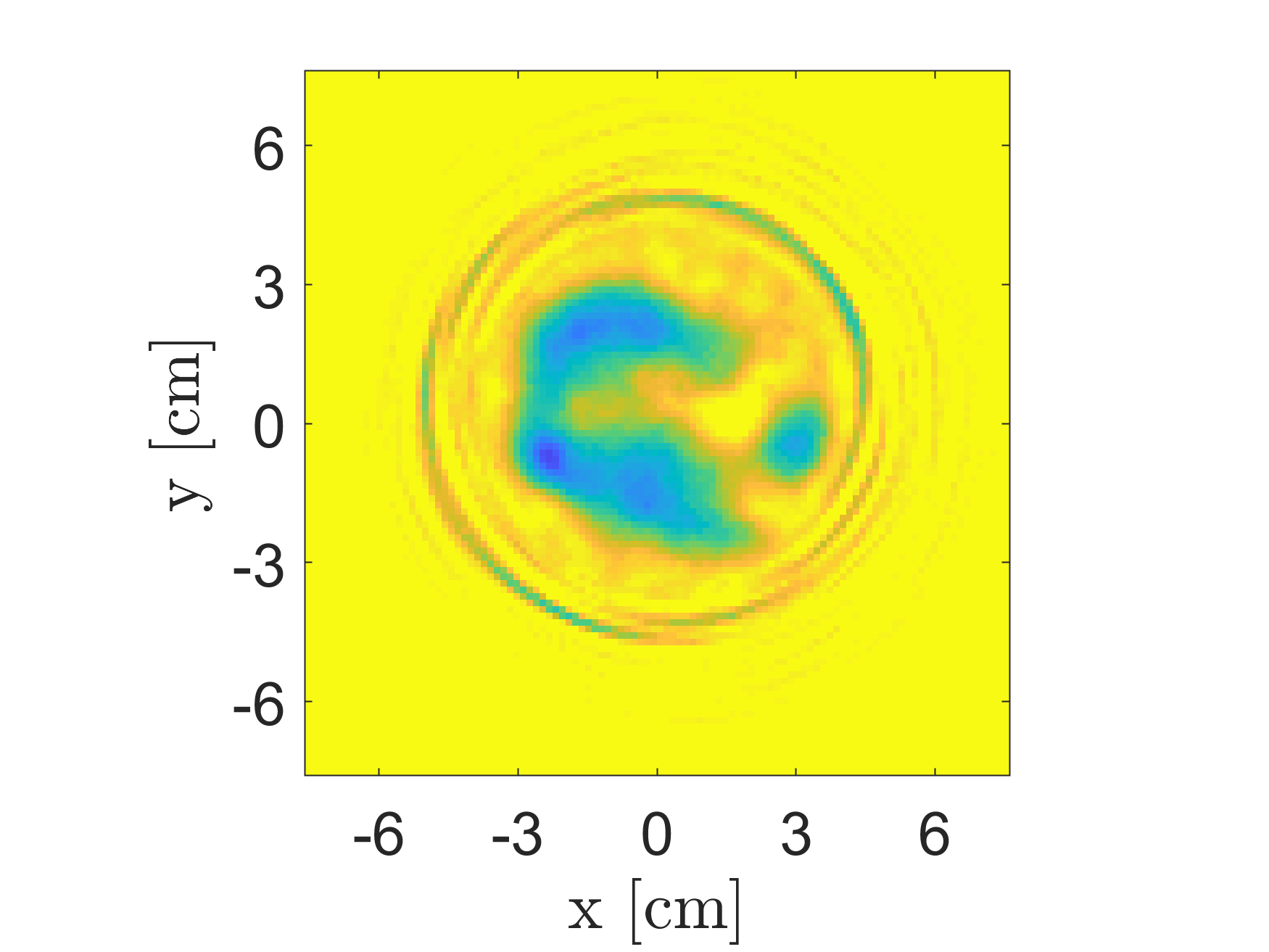}} &
        {\includegraphics[draft=false, trim=2cm 1cm 1.6cm 0cm, clip, height=3.7cm]{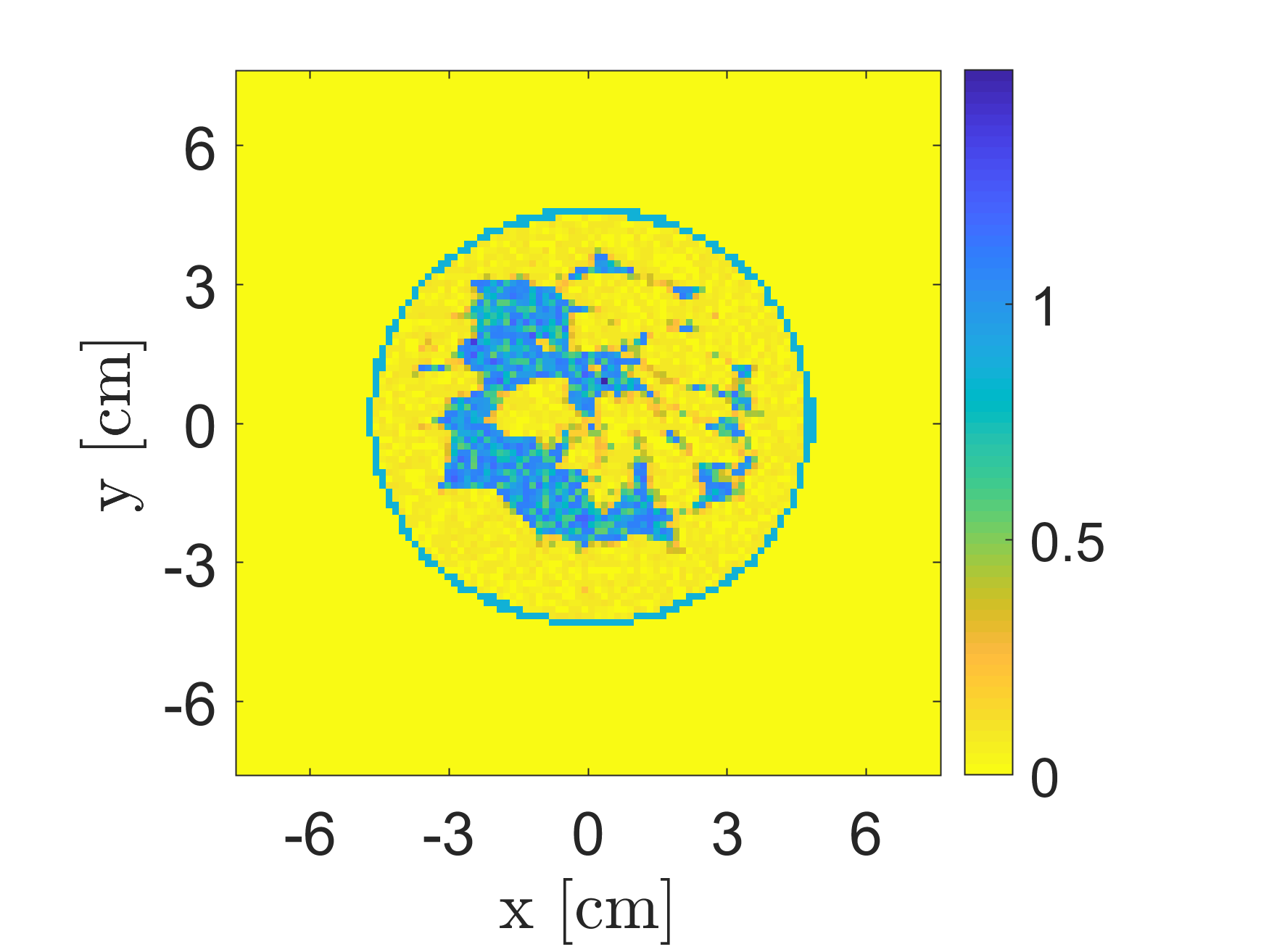}}  \\
(e) & (f) & (g) & (h) \\
        {\includegraphics[draft=false, trim=2.6cm 1cm 1.6cm 0cm, clip, height=3.7cm]{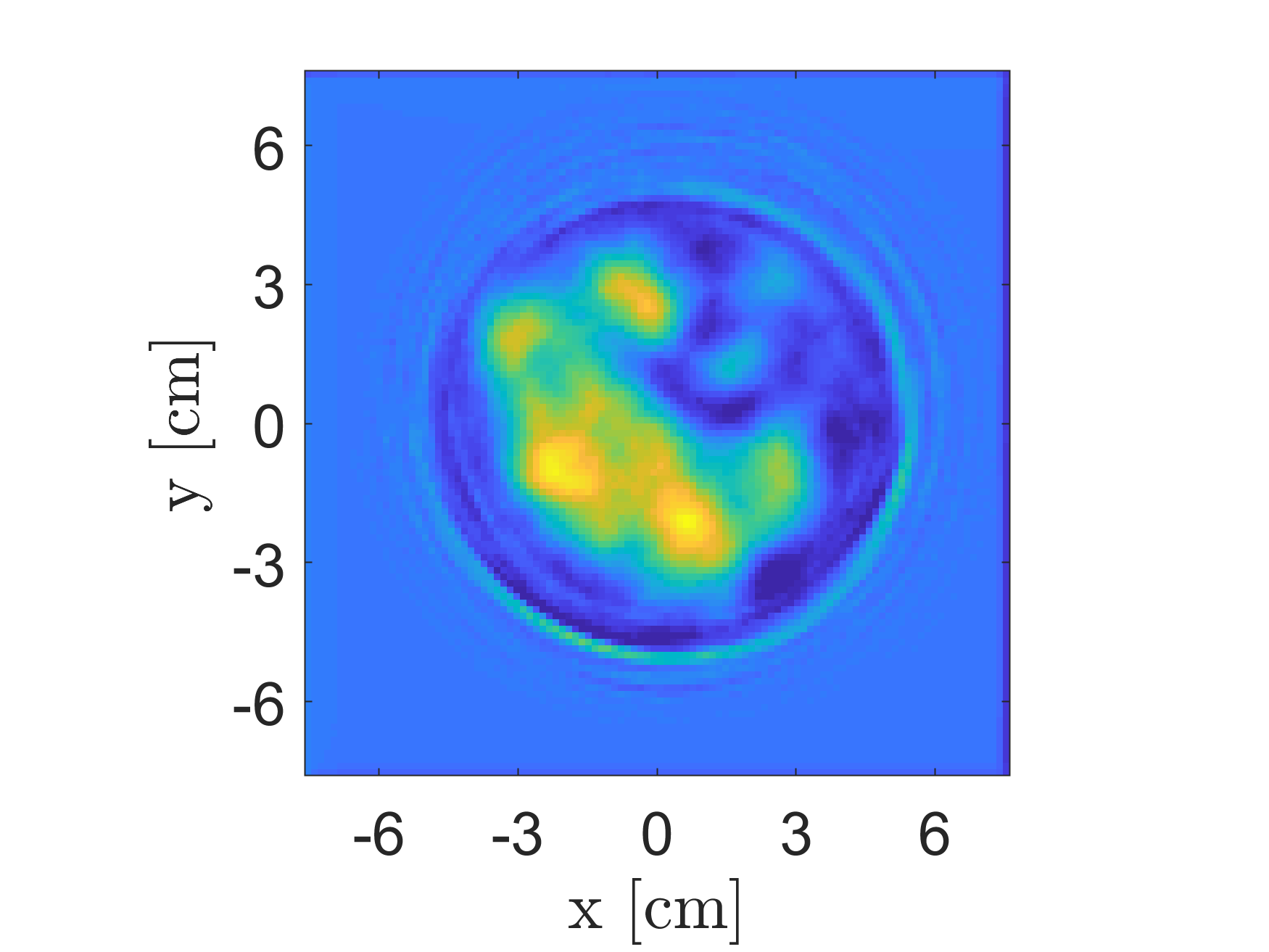}} & 
        {\includegraphics[draft=false, trim=2cm 1cm 1.6cm 0cm, clip, height=3.7cm]{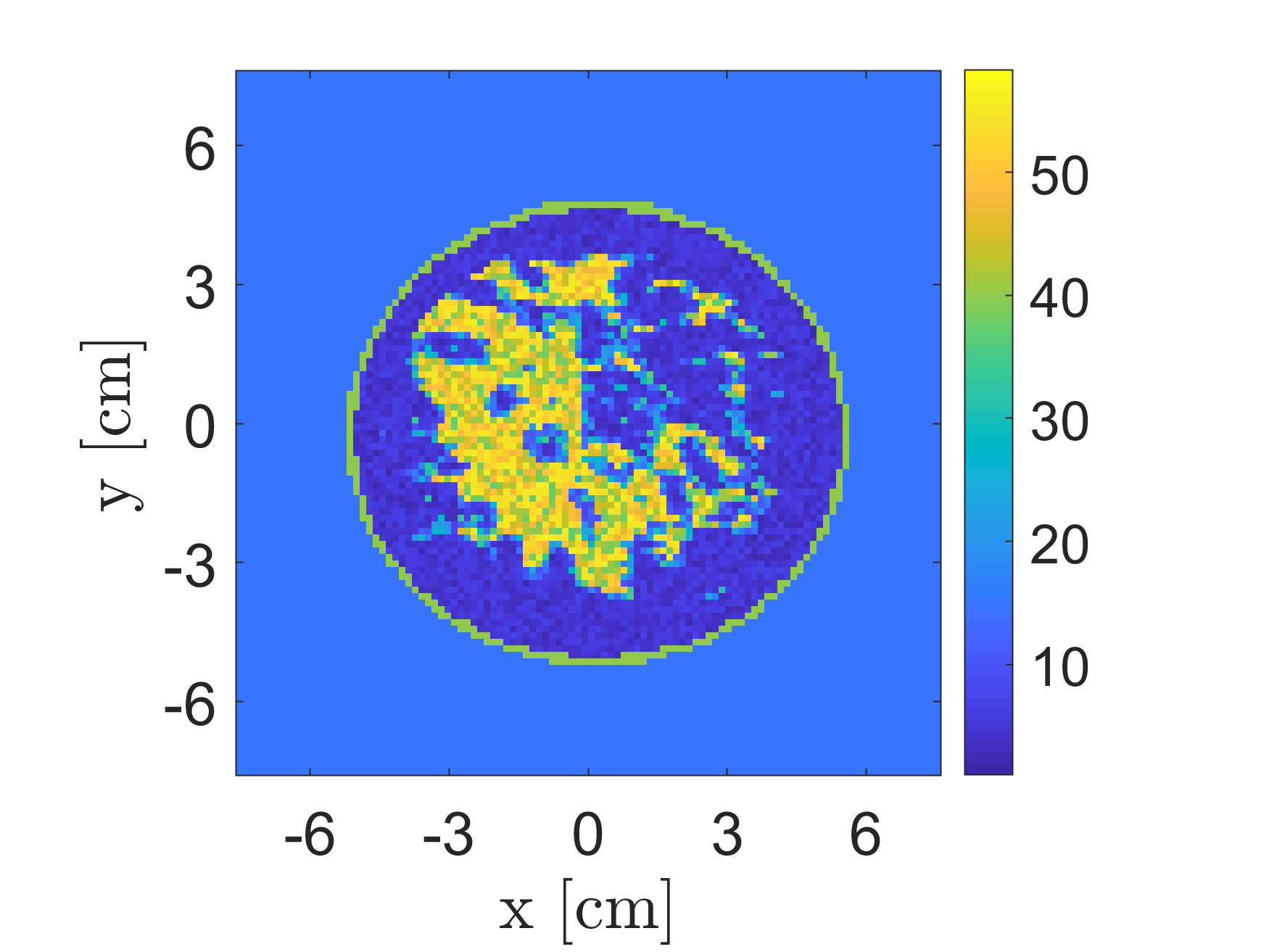}} &
        {\includegraphics[draft=false, trim=2.6cm 1cm 1.6cm 0cm, clip, height=3.7cm]{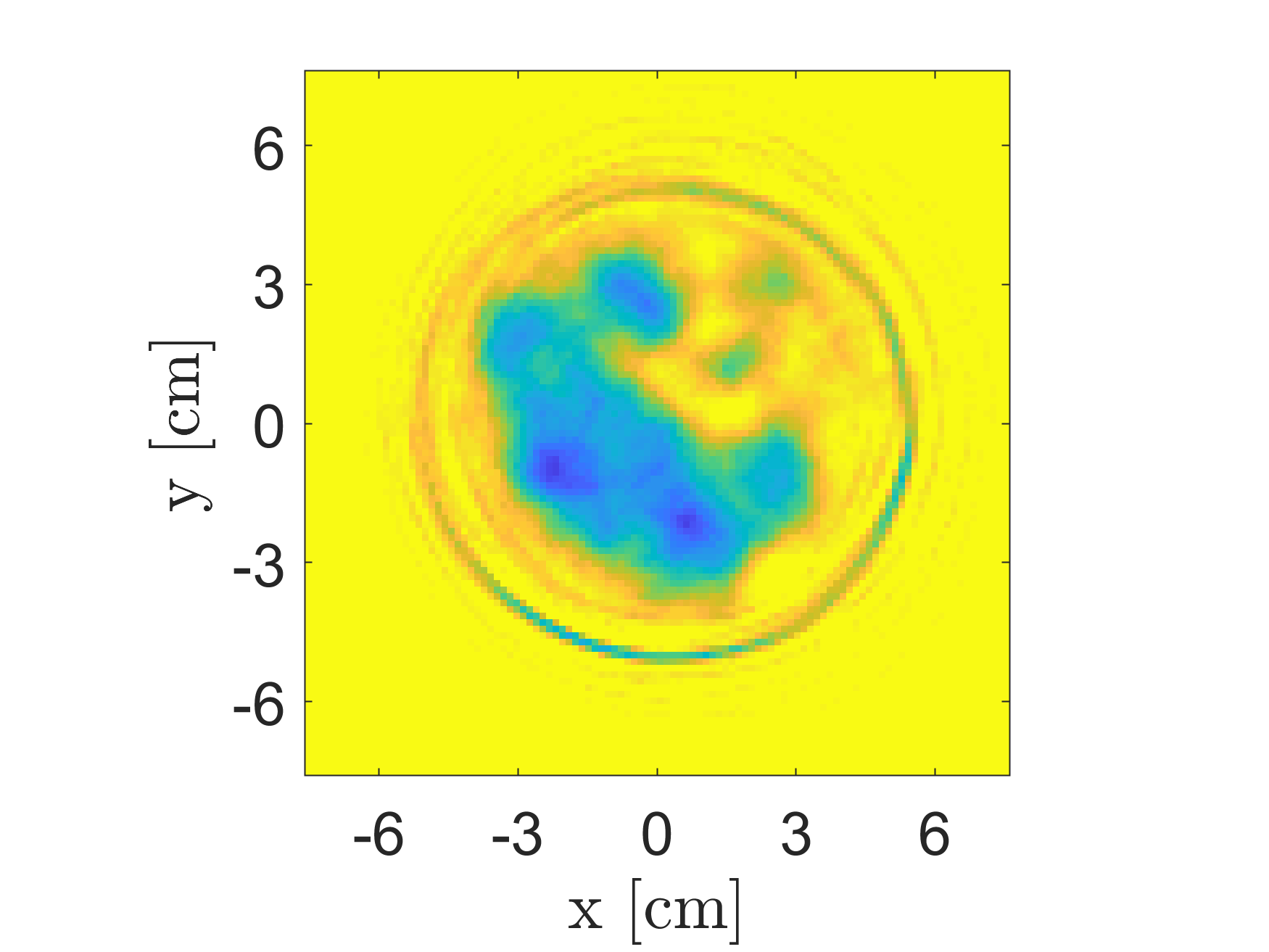}} &
        {\includegraphics[draft=false, trim=2cm 1cm 1.6cm 0cm, clip, height=3.7cm]{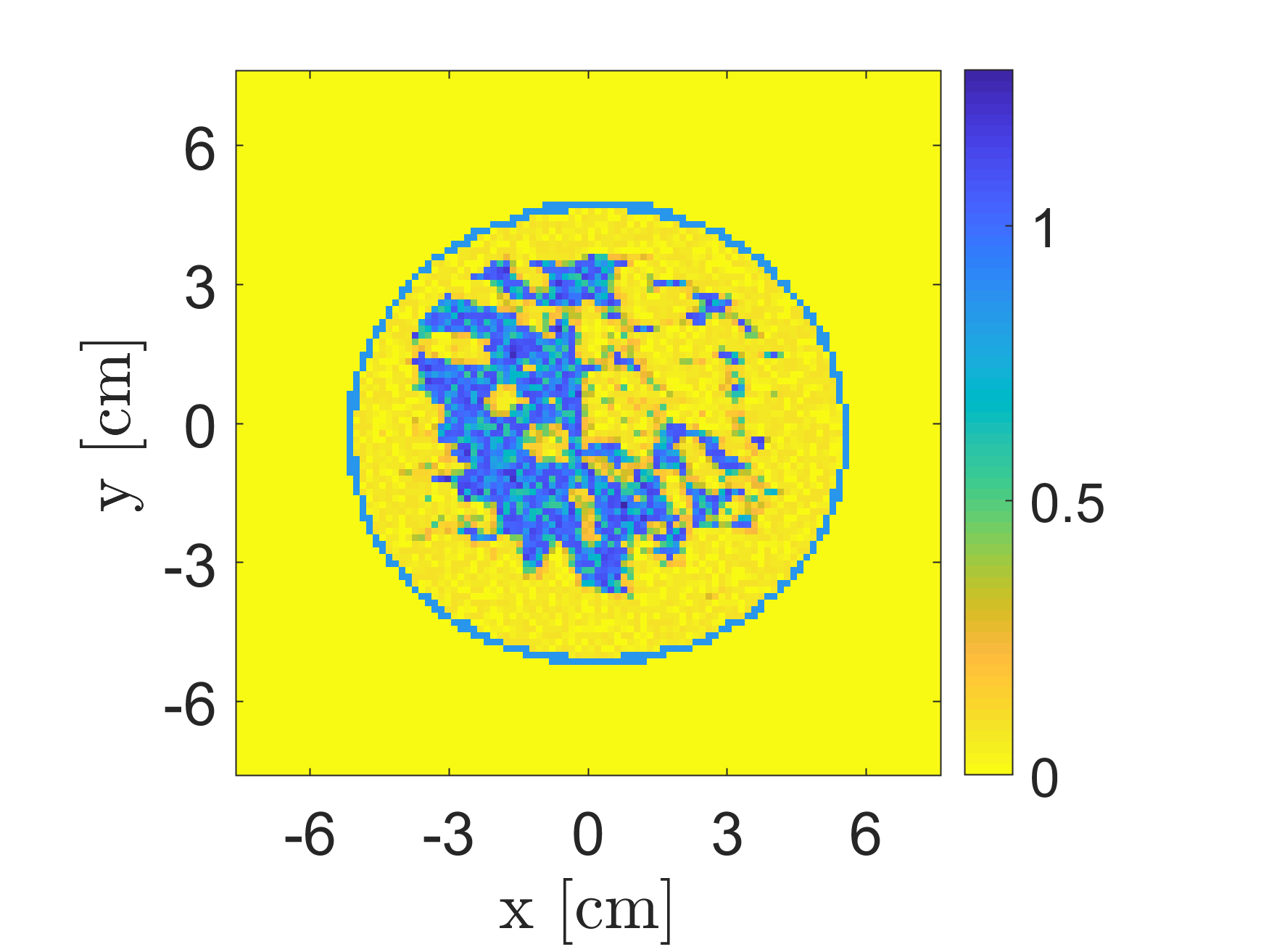}}  \\
(i) & (j) & (k) & (l) \\
   \end{tabular}}
   \caption{Recoveries obtained via the proposed ANN-based approach for some realistic breast phantoms proposed by \cite{lazebnik2007large}. Class 2 (scattered fibroglandular), breast ID: 012204, slice 54: retrieved permittivity and conductivity maps (a, c) and reference (b, d). Class 3 (heterogeneously dense), breast ID: 062204, slices 47 and 71: retrieved permittivity and conductivity maps (e, i, g, k), and references (f, j, h, l), respectively.}  
   \label{fig:lazebnik_inversion}
\end{figure*}

\subsection{Comparison with other Nonlinear Approaches}
The performance of the almost real-time ANN-based approach proposed previously has been compared with other well-known nonlinear methods. 
More in detail, the comparison was performed with the distorted Born iterative method (DBIM) \cite{chew1990reconstruction, chew1995waves} and the contrast source inversion (CSI) in their standard formulation \cite{van2001contrast}. 
In the former case, a solution to the considered inverse problem is sought via a series of linear problems which gradually recover the full nonlinear profile, while in the latter case an iterative inversion scheme based on a functional minimisation involving both Eqs. \eqref{eq:ES_equations_data}-\eqref{eq:ES_equations_domain} is exploited.
These conventional nonlinear inversion approaches are iterative and usually employ local minimisation schemes in order to avoid prohibitive computational time required by global optimisation schemes. Under these assumptions, the choice of the initial guess for starting the minimisation procedure becomes paramount for obtaining good recoveries.

Figs. \ref{fig:net_comparisons_re}-\ref{fig:net_comparisons_im} show the retrieved complex permittivity maps (i.e., real and imaginary parts) for three different breast profiles belonging to classes II and III at the operating frequency of 1 GHz (i.e., a wavelength of approximately 3.5 cm in case of fibro-glandular tissues) and with data corrupted by additive white Gaussian noise (AWGN) with a signal-to-noise ratio (SNR) equal to 30 dB. Further information regarding the considered configuration and the data generation is reported in Subsection \ref{sec:methodology:phantom_gen}.    
More specifically, these figures compare the recoveries obtained by the proposed ANN approach with other conventional nonlinear inversion strategies, i.e. the DBIM and the CSI. It is worth to note that the recoveries via DBIM and CSI were obtained starting the local minimisation procedure from an initial guess equal to the background medium. By comparing the recoveries it is clear that the ANN approach outperforms these classical techniques, providing good performance also on conductivity estimate, which represents the hardest part to be retrieved by conventional methods and also the most important information for diagnostic and therapeutic treatments. 

Concerning the computational time, the proposed ANN-based approach results in an almost real-time procedure, conversely from conventional nonlinear approaches which are time-consuming per each frequency and, sometimes, also per each iteration (e.g., the DBIM case, for which a forward solver has to be run at each inversion iteration, involving at least 5-hour simulation for 15 iterations on the same workstation).
It is worth to observe that the proposed ANN approach is able to provide information about the shape and location of the breast and, more specifically, about the skin layer, which still represents an issue for its dielectric features and thickness at the considered operating frequency. 

Furthermore, the good recovery performance noticeable in the reconstructions related to tissues relative permittivity maps are also evident in the corresponding conductivity, which still represents the hardest, most challenging part in nowadays inverse scattering imaging approaches.   

In order to test the robustness of the proposed method versus noise, Fig. \ref{fig:perf_ass_VS_noise} illustrates a comparison among the recoveries of the selected breast phantoms reported in Figs. \ref{fig:net_comparisons_re}- \ref{fig:net_comparisons_im} for a much lower value of the SNR, i.e. 5 dB. These images prove that the proposed approach is robust against noise and has good inversion performance also in very noisy scenarios. 

Finally, to prove further the robustness of the approach, its recovery performance was tested on realistic numerical breast phantoms generated from magnetic resonance images of female breasts collected by the Cross-disciplinary Electromagnetics Laboratory at University of Wisconsin-Madison (CEM-UW) \cite{lazebnik2007large}. 
More in detail, some slices of the breast phantoms reported in the repository were considered, i.e. class 2 (scattered fibroglandular), breast ID: 012204, slice 54, and class 3 (heterogeneously dense), breast ID: 062204, slices 47 and 71. The measurement configuration and frequency were the same of the ones reported in Subsection \ref{sec:methodology:phantom_gen} and an AWGN noise with SNR = 30 dB was applied on the data. 
As proved by the results shown in Fig. \ref{fig:lazebnik_inversion}, the accuracy of the recoveries is still good allowing better retrieving performance especially on the conductivity of the considered breast profiles, but also a better resolution of the skin layer and a much faster inversion rather than conventional approaches.

\section{Conclusion}
In this paper a fully-connected artificial neural network for almost real-time microwave breast imaging applications has been proposed. The considered network was trained by exploiting an in-house realistic breast-like phantom generator adopted for the data set population and by in-house forward solver codes for the generation of the scattered field data matrices to be exploited in the training of the network. 

The methodology was tested in noisy scenarios with different values of SNR proving good robustness against noise (i.e., 30 and 5 dB) and a performance assessment for the choice of the network architecture was carried out.
It is worth to note that the recovery performance depends on the network architecture and each topology provides an improvement in retrieving information in proper bandwidths of the scenario under test.

The results are very promising in comparison with conventional nonlinear inverse scattering approaches from a quantitative point of vies as well as for the computation burden. Conversely from standard approaches, the proposed methodology is almost real-time and allows to obtain better recoveries, especially for the conductivity map which represents the most difficult part in inverse scattering approaches as well as the most useful information for biomedical applications in diagnostics and therapy.

Future work will focus on the design of advanced neural network strategies to improve the quality of the results and to process three-dimensional data set for the detection and characterisation of potential malignant areas. Furthermore, the combination of the proposed approach with convolutional neural networks and other deep learning strategies will be also explored.

\end{document}